\documentclass[usegraphicx]{mn2e}
\usepackage{graphicx}
\usepackage{rotating}
\usepackage{amsmath}

\def\lesssim{\mathrel{\hbox{\rlap{\hbox{\lower4pt\hbox{$\sim$}}}\hbox{$<$}}}}

\begin{document}
\title[An explanation for the AGN soft excess]{An explanation for the soft X-ray excess in AGN}
   \author[J. Crummy et al.]{J. Crummy$^1$\thanks{E-mail: jc@ast.cam.ac.uk}, A.C. Fabian$^1$, L. Gallo$^2$ and R.R. Ross$^3$\\
   $^1$Institute of Astronomy, Madingley Road, Cambridge, CB3 0HA\\
   $^2$Max-Planck-Institut f\"ur extraterrestrische Physik, Postfach 1312, 85741 Garching, Germany\\
   $^3$Physics Department, College of the Holy Cross, Worcester, MA 01610, USA}
\maketitle

\begin{abstract}
We present a large sample of type 1 active galactic nuclei (AGN) spectra taken with \textit{XMM-Newton}, and fit them with both the conventional model (a power law and black body) and the relativistically-blurred photoionized disc reflection model of Ross \& Fabian (2005). We find the disc reflection model is a better fit. The disc reflection model successfully reproduces the continuum shape, including the soft excess, of all the sources. The model also reproduces many features that would conventionally be interpreted as absorption edges. We are able to use the model to infer the properties of the sources, specifically that the majority of black holes in the sample are strongly rotating, and that there is a deficit in sources with an inclination $> 70^{\circ}$. We conclude that the disc reflection model is an important tool in the study of AGN X-ray spectra.
\end{abstract}

\begin{keywords}
 accretion, accretion discs -- galaxies: active -- X-rays: galaxies.
\end{keywords}

\section{Introduction}
\begin{figure*}
\begin{center}
\includegraphics[angle=270, width=0.95\textwidth]{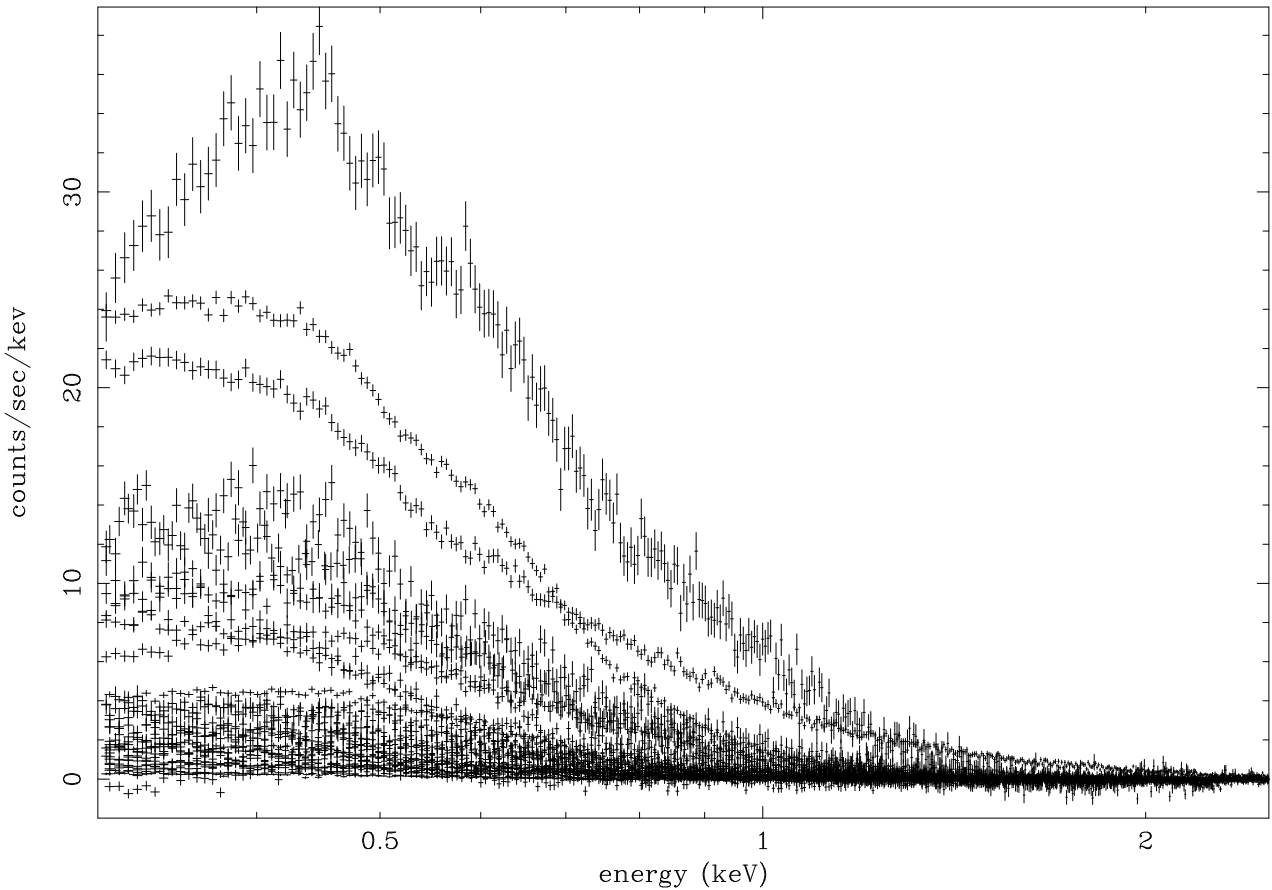}
\caption{The soft excesses of all the sources in our sample. The sources were fit as in Table \ref{phenom_fit_table}, then the black body component was removed and the residuals were plotted. Absorption edges were left in the model to show the form of the soft excess more clearly. The top three sources, in descending order of flux at 0.5~keV, are ARK 564, NGC 4051 and TON S180; note the unusually extended shape of the TON S180 soft excess.}
\label{soft_excess_figure}
\end{center}
\end{figure*}

\begin{figure}
\begin{center}
\includegraphics[angle=270, width=0.45\textwidth]{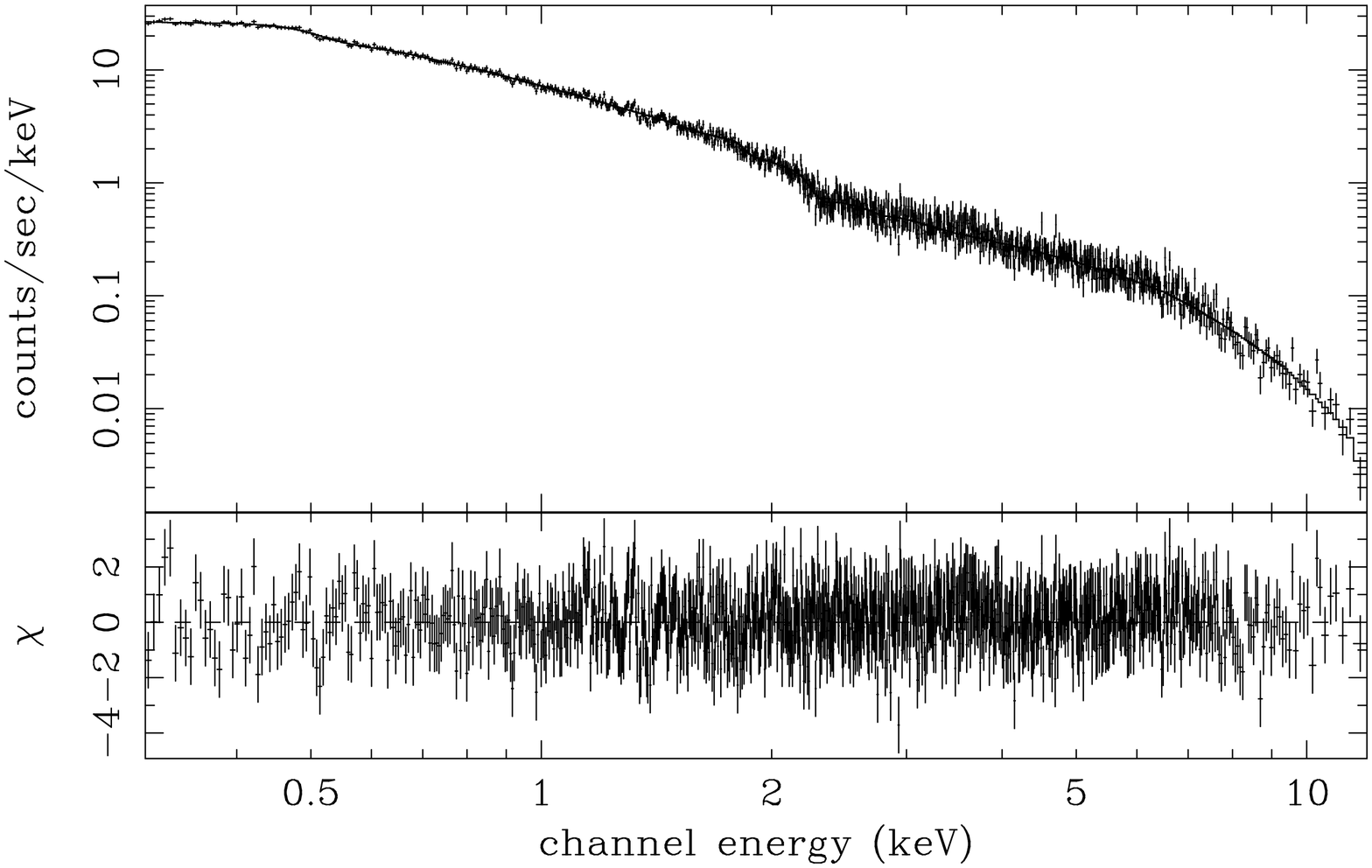}
\includegraphics[angle=270, width=0.45\textwidth]{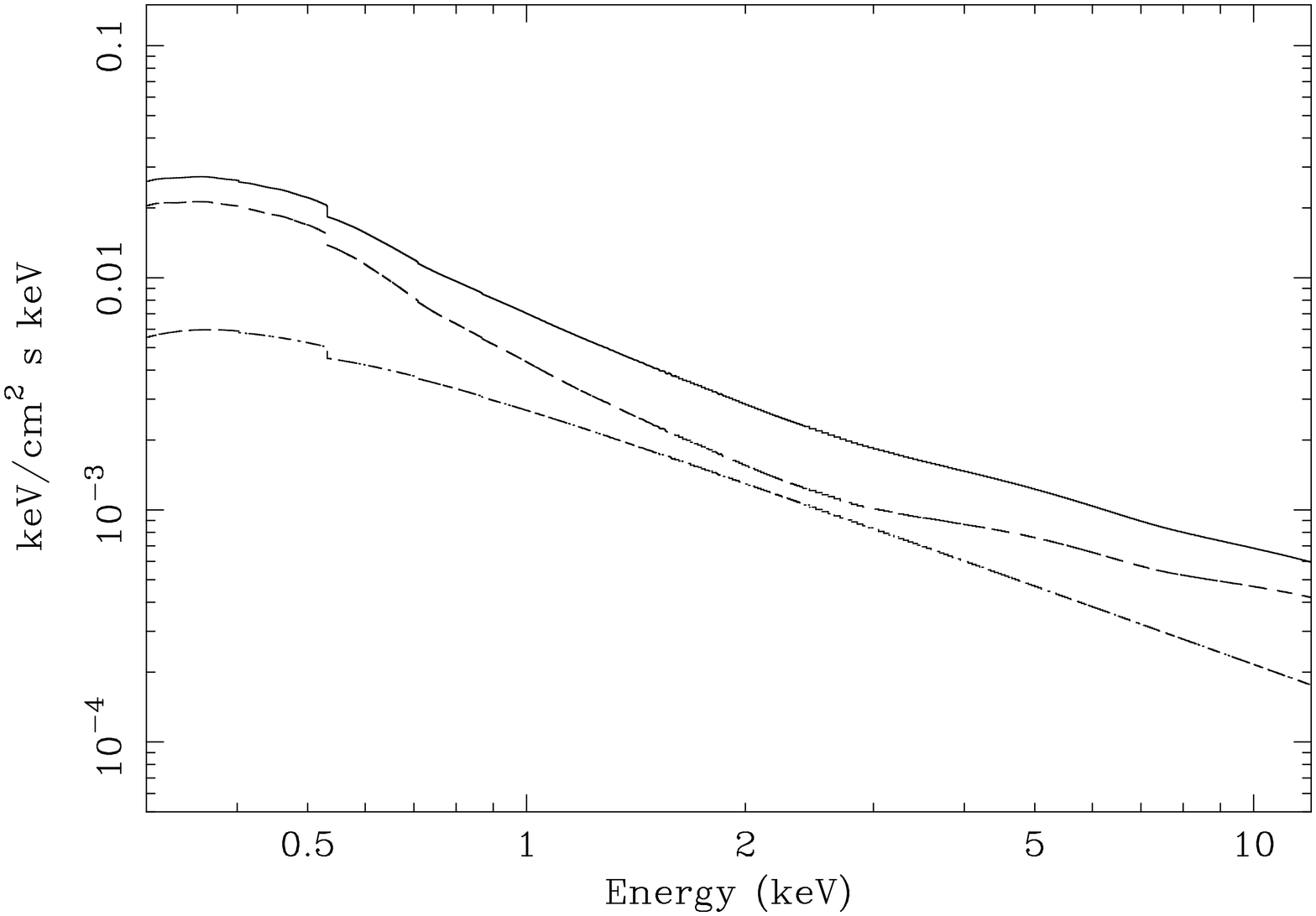}
\caption{The relativistically-blurred photoionized disc reflection model fit and residuals to PG 1501+106 (top) and a plot of the model used showing the illuminating power law and reflection components (bottom). The dashed line is the reflection component, the dash-dotted line is the power law, and the solid line is the total model. The model is shown over an extended energy range.  The curvature at low energies is due to cold absorption local to our Galaxy. See Table \ref{disc_fit_table} for fit parameters.}
\label{1501_spectrum_figure}
\end{center}
\end{figure}

Active galactic nuclei are classified into two types: type 1 AGN have both broad and narrow components of their optical emission lines, type 2 AGN show only narrow lines. The narrow lines are emitted from the outer regions of the source, the broad lines from matter in the inner regions where Doppler broadening increases the width of the lines. The unified model of AGN explains the two types as an inclination effect; Seyfert 2 galaxies are observed at high inclinations and have the central regions obscured by a torus, Seyfert 1 galaxies are seen at low inclinations where the line of sight passes over the torus. Therefore, we expect that our sample of type 1 AGN will allow us to investigate the central regions of these systems.\\
The soft excess is a major component of the spectra of many AGN in the X-ray band and is present in every source in this survey, see Figure \ref{soft_excess_figure}. First noted in Arnaud et al. (1985), it is extra emission below $\sim$1.5~keV than expected from extrapolating the power law spectrum observed at higher energies. The soft excess is usually well fit with a black body which has a roughly constant temperature of 0.1 -- 0.2~keV over several decades in AGN mass (Walter \& Fink 1993, Czerny et al. 2003, Gierli\'{n}ski \& Done 2004). If it is thermal, this temperature is much too high to be explained by the standard accretion disc model of Shakura \& Sunyaev (1973), although it could be explained by a slim accretion disc in which the temperature is raised by photon trapping (e.g. Abramowicz et al. 1998; Mineshige et al. 2000) in which case the accretion is super-Eddington (Tanaka, Boller \& Gallo 2005), or by Comptonisation of EUV accretion disc photons (e.g. Porquet et al. 2004). Atomic physics could trivially explain the constant temperature of the component, e.g. strong, relativistically-blurred absorption from a disc wind (Gierli\'{n}ski \& Done 2004) or photoionized emission blurred relativistically by motion in an accretion disc (Ballantyne, Iwasawa \& Fabian 2001; this paper).\\
Relativistically-blurred photoionized iron emission has been observed in several AGN: a clear relativistic profile in a few sources (e.g. MCG -6-30-15, Tanaka et al. 1995; Vaughan \& Fabian 2003), and extended red wings to iron lines in many sources (e.g. Nandra et al. 1997). This concept is extended in disc reflection models which reproduce the emission expected from a photoionized accretion disc around a black hole (e.g. Ross \& Fabian 1993; Ballantyne, Ross \& Fabian 2001), models which have been used with some success in e.g. Ballantyne, Iwasawa \& Fabian (2001), Fabian et al. (2002), Ballantyne, Vaughan \& Fabian (2003). The work presented here is the largest study to date, and is focused on fitting many objects with a uniform analysis rather than a detailed study of one or a few sources. We uses the latest model from Ross \& Fabian (2005), in which a semi-infinite slab of optically thick cold gas of constant density is illuminated by a power law, producing a Compton component and fluorescence lines from the ionized species in the gas, the illuminating and reflected components are added together (with the amount of each free to vary) and the total emission convolved with a Laor (1991) line profile to simulate the blurring from a relativistic accretion disc. The Laor profile assumes a flat accretion disc with a power law emissivity and sharp edges at the inner and outer radius around a maximally-rotating black hole. The Laor profile has as free parameters the emissivity index, the inner and outer radius (in gravitational radii; the blurring process does not depend on black hole mass) and the inclination of the disc to the line of sight. The reflection spectra allow fitting to the iron abundance of the gas, the spectral index of the illuminating power law, and the ionization parameter (a measure of the ratio of the energy density in the illuminating radiation to the atomic number density in the illuminated gas). Examples of the model are given in Figs. \ref{1501_spectrum_figure} and \ref{1202_emo_figure}. This model is physically motived, and therefore may prove useful in understanding these sources.\\

\section{The sources}
In this paper we use publicly available archival \textit{XMM-Newton} data on 22 type 1 AGN from the Palomar-Green (PG) sample and a selection of 12 other Seyfert 1 galaxies with high-quality observations available. Where multiple observations are publicly available we use the longest where the \texttt{pn} took data. A list of the sources with their properties and the observation IDs used is given in Table \ref{sources_table}. Most of these sources are the subject of many papers and have been observed with several different X-ray instruments, however, we confine ourselves to \textit{XMM-Newton} data and perform a standard analysis across all of them.\\

\begin{table*}
\caption{List of sources investigated. The Galactic absorption column is taken from the \textsc{nh} ftool (Dickey \& Lockman 1990).}
\label{sources_table}
\begin{tabular}{lllccc}
\hline
Source & Alternate name & redshift & Galactic absorption & Mass & Observation ID\\
 & & & column ($\times 10^{20}$~cm$^{-2}$) & $\log(M/M_{\odot})$&  \\
\hline
PG 0003+199 & MRK 0335 & 0.025785 & 3.99& 7.07$^{a}$ & 0101040101\\
PG 0050+124 & I Zw 001 &  0.061142 & 4.99 & 7.13$^{a}$ &0110890301\\
PG 0157+001 & MRK 1014 & 0.163009 & 2.59 & 8.00$^{a}$ &0101640201\\
PG 0844+349 & & 0.064000 & 2.98 & 7.66$^{a}$ &0103660201\\
PG 0947+396 & & 0.206 & 1.57 & 8.46$^{a}$ &0111290101\\
PG 0953+414 & & 0.2341 & 1.14& 8.52$^{a}$ &0111290201 \\
PG 1048+342 & & 0.167 & 1.83& 8.14$^{a}$ &0109080701 \\
PG 1115+407 & & 0.154 & 1.91& 7.44$^{a}$ &0111290301\\
PG 1116+215 & TON 1388 & 0.1765 & 1.28& 8.41$^{a}$ &0111290401\\
PG 1202+281 & GQ Com & 0.1653 & 1.67& 8.37$^{a}$ &0109080101\\
PG 1211+143 & & 0.0809 & 2.75& 7.81$^{a}$  &0112610101\\
PG 1244+026 & & 0.048 & 1.75& 6.24$^{a}$ &0051760101 \\
PG 1307+085 & & 0.155 & 2.11 & 8.50$^{a}$  &0110950401\\
PG 1309+355 & TON 1565 & 0.184 & 1.03 & 8.20$^{a}$ &0109080201\\
PG 1322+659 & & 0.168 & 2.01& 7.74$^{a}$ & 0109080301\\
PG 1352+183 & E 1352+183 & 0.152 & 2.05& 8.20$^{a}$ & 0109080401 \\
PG 1402+261 & TON 0182 & 0.164 & 1.47& 7.76$^{a}$ & 0109081001\\
PG 1404+226 & & 0.098 & 2.14 & 6.65$^{a}$ & 0051760201\\
PG 1427+480 & [HB89] 1427+480 & 0.221 & 1.88& 7.86$^{a}$ & 0109080901\\
PG 1440+356 & MRK 0478 & 0.079055 & 1.03& 7.28$^{a}$ & 0107660201\\
PG 1444+407 & [HB89] 1444+407 & 0.267300 & 1.25& 8.17$^{a}$ & 0109080601\\
PG 1501+106 & MRK 0841 & 0.036422 & 2.34 & 8.23$^{a}$ & 0070740301\\
NGC 4051 & 1WGA J1202.1+444 & 0.002336 & 1.32 & 6.13$^{b}$ & 0109141401\\
IRAS 13349+2438 & [HB89] 1334+246 & 0.107641 & 1.16& & 0096010101\\
ARK 564 &  1H 2239+294 & 0.024684 & 6.40& 6.46$^{c}$ & 0006810101\\
MRK 1044 & MCG -02-07-024 & 0.016451 & 3.55& 6.23$^{c}$ & 0112600301\\
E 1346+266 & 1WGA J1348+2622 & 0.915 & 1.18& & 0109070201\\
MRK 0359 & 1WGA J0127.5+1910 & 0.017385 & 4.80& 6.23$^{c}$ & 0112600601\\
PHL 1092 & [HB89] 0137+060 & 0.396 & 4.07& & 0110890901\\
RE J1034+396 & & 0.42144 & 1.47& & 0109070101\\
PKS 0558-504 & [HB89] 0558-504 & 0.137 & 4.38& & 0125110101\\
MRK 0766 & RX J1218.4+2948 & 0.012929 & 1.71 & 6.63$^{c}$ & 0096020101 \\
MRK 0586 & NAB 0205+02 & 0.155300 & 3.51 & 7.86$^{b}$ & 0048740101\\
TON S180 & 1WGA J0057-2222  & 0.061980 & 1.55& 7.06$^{c}$ & 0110890401\\
\hline
\end{tabular}
\\
${}^{a}$ Mass taken from Gierli\'{n}ski \& Done (2004), who derive them from Boroson (2002). ${}^{b}$ Mass from Woo \& Urry (2002). ${}^{c}$ Mass from Wang \& Lu (2001).
\end{table*}

\section{Data reduction}
We obtained the Observation Data Files (ODFs) from the \textit{XMM-Newton} public archive and reduced them in the standard way using \textsc{SAS 6.0} to produce event lists. We extracted spectra and light-curves using circular source-centred regions 40 arcsec in size for the \texttt{pn} and 60 arcsec in size for the \texttt{MOS}, using a smaller extraction region where appropriate due to chip gaps, etc. We used the standard valid event patterns of 0 -- 4 (singles and doubles) for \texttt{pn} and 0 -- 12 (singles to quadruples) for the \texttt{MOS}. Background spectra and light-curves were similarly created using regions away from any sources, possible out-of-time event trails, and chip gaps. Where the count rate of background flares was greater than 5 per cent of the source count rate a Good Time Interval (GTI) file was used to exclude those events, and even more conservative background filtering was adopted where possible. We used the \textsc{SAS} tasks \textsc{rmfgen} and \textsc{arfgen} to create the response matrices.\\
As the \texttt{MOS} cameras are not accurately calibrated against the \texttt{pn} (Kirsch et al. 2004) only one of the two systems can be used without introducing complications to the fits. We chose to use \texttt{pn} data as it has a larger effective area and spectral range. We did not reduce \texttt{RGS} or \texttt{OM} data.\\
The \textsc{SAS} task \textsc{epatplot} was used on the extracted spectra to test for the presence of pile up. All but five of the observations were found to be pile-up free. The piled up observations were re-extracted using an annular region to exclude the piled-up centre of the observation. Annulus sizes used were 11, 8, 8, 16 and 5 arcsec for PG 0003+199, PG 0844+349, PG 1244+026, PG1440+356 and RE J1034+396, respectively.\\
We performed the spectral analysis using \textsc{xspec v11.3} (Arnaud 1996), taking the limits in accurate calibration of the \texttt{pn} data as 0.3 -- 12.0~keV. We grouped the spectra so each bin would include 20 or more source counts so that $\chi^{2}$ statistics would be applicable. All quoted errors are 90 per cent limits on one parameter ($\Delta \chi^{2}$ = 2.706).\\
We produced background-subtracted light-curves and hardness ratios for the sources. We found that 12 of the 34 sources display some variability in their hardness ratios over the course of the observations. In the interest of using the highest quality data available, we analysed only spectra integrated over the entire observation. This strategy may have introduced some error, but a detailed analysis of spectral variability of a large number of sources is beyond the scope of this paper.\\
\section{Analysis and results}\label{results}
\begin{figure}
\begin{center}\includegraphics[width=0.45\textwidth]{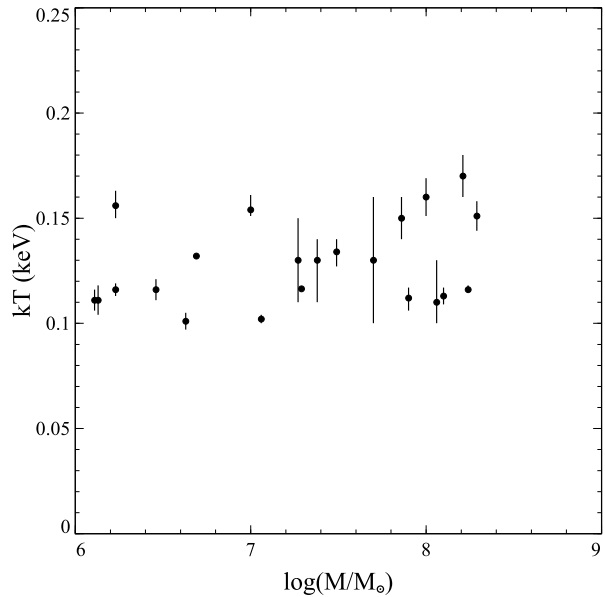}
\caption{Soft excess black body temperature ($kT$) measured for the simple model plotted against black hole mass ($M$), see Section \ref{results} and Table \ref{sources_table}. The soft excess is confined to a narrow temperature band over several decades in mass, as earlier noted by Gierli\'{n}ski \& Done (2004) and others.}
\label{soft_temp_figure}
\end{center}
\end{figure}

\begin{figure*}
\begin{center}
\includegraphics[angle=270, width=0.4\textwidth]{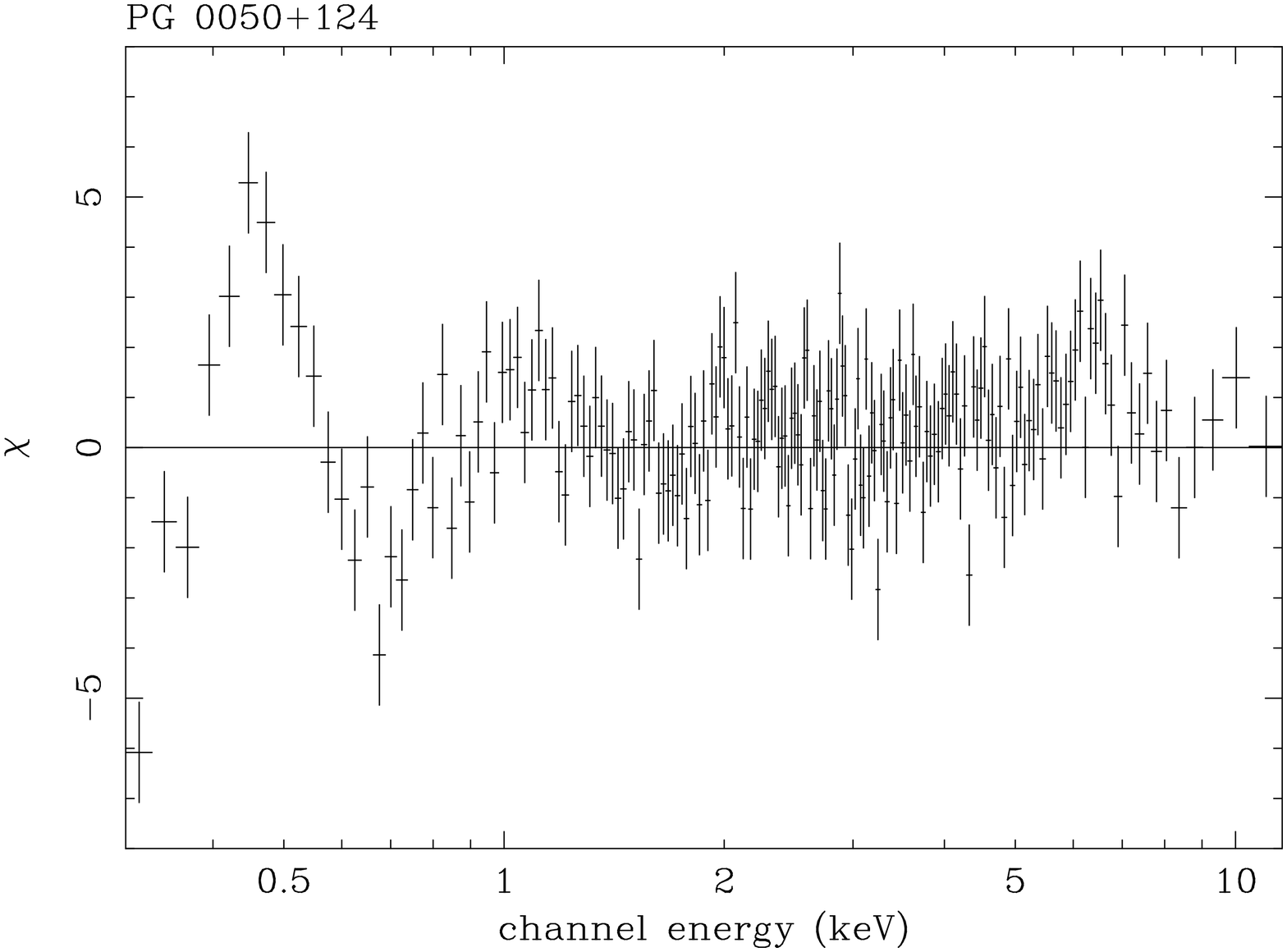}
\hspace{11mm}
\includegraphics[angle=270, width=0.4\textwidth]{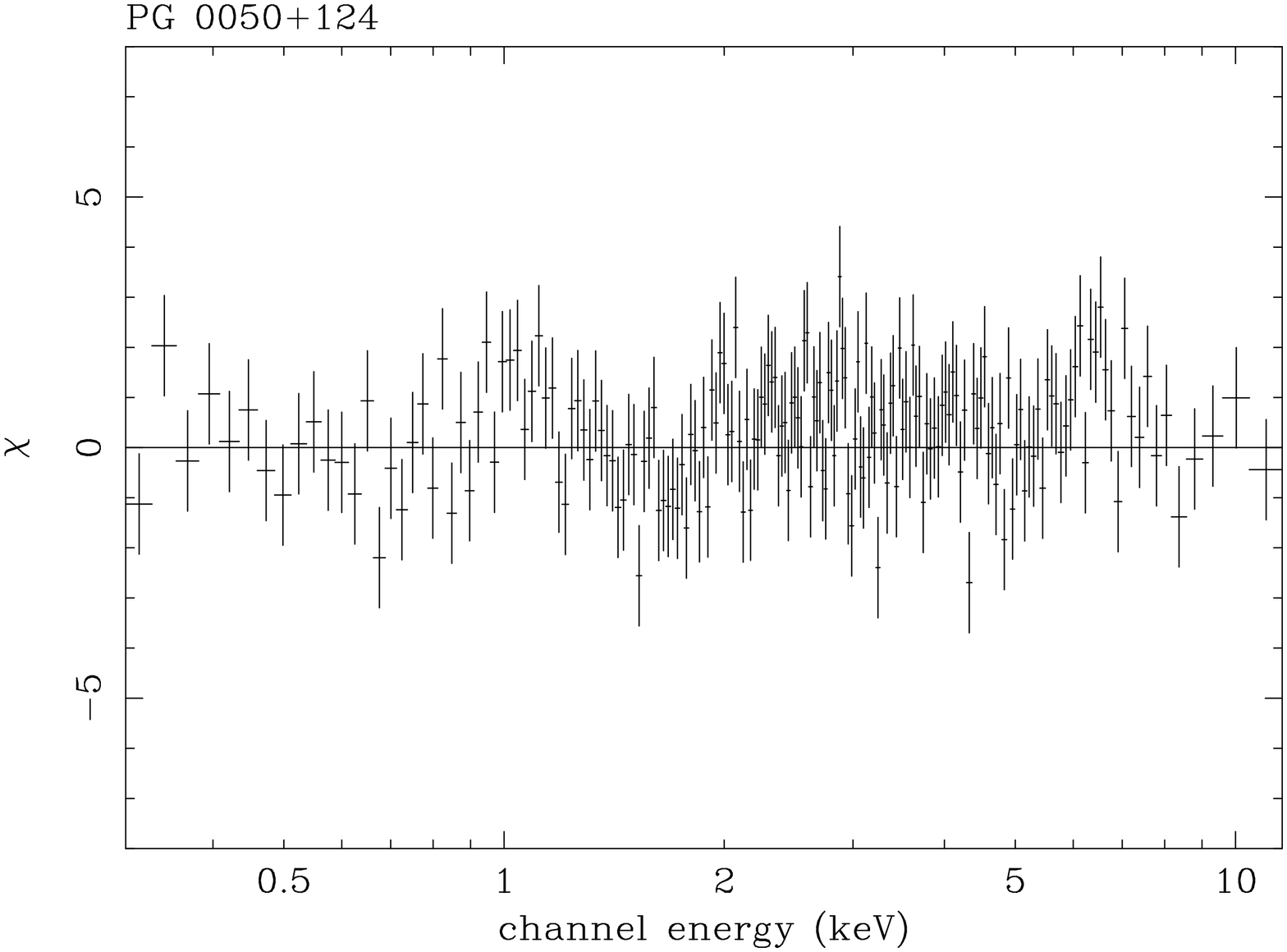}
\includegraphics[angle=270, width=0.4\textwidth]{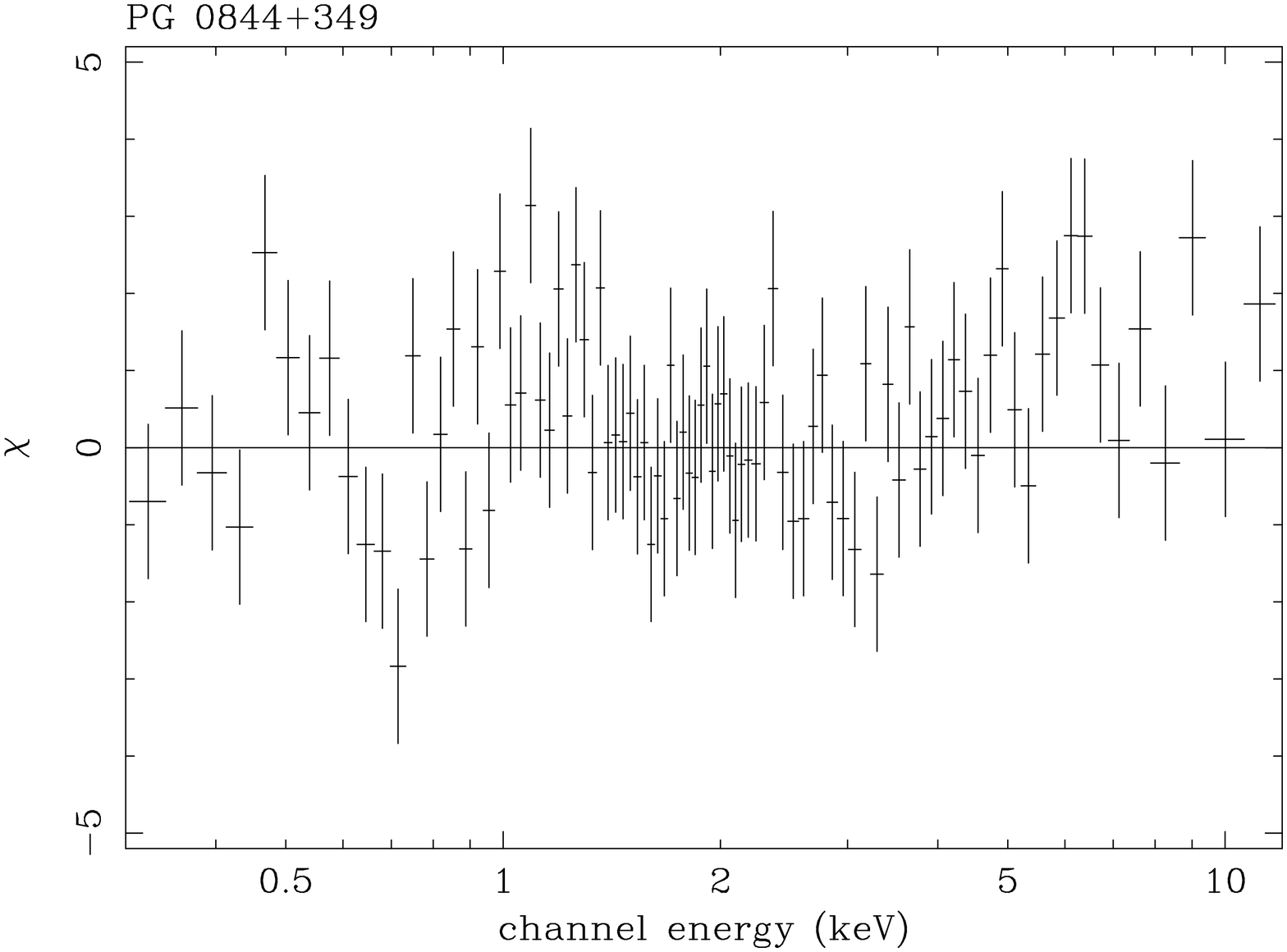}
\hspace{11mm}
\includegraphics[angle=270, width=0.4\textwidth]{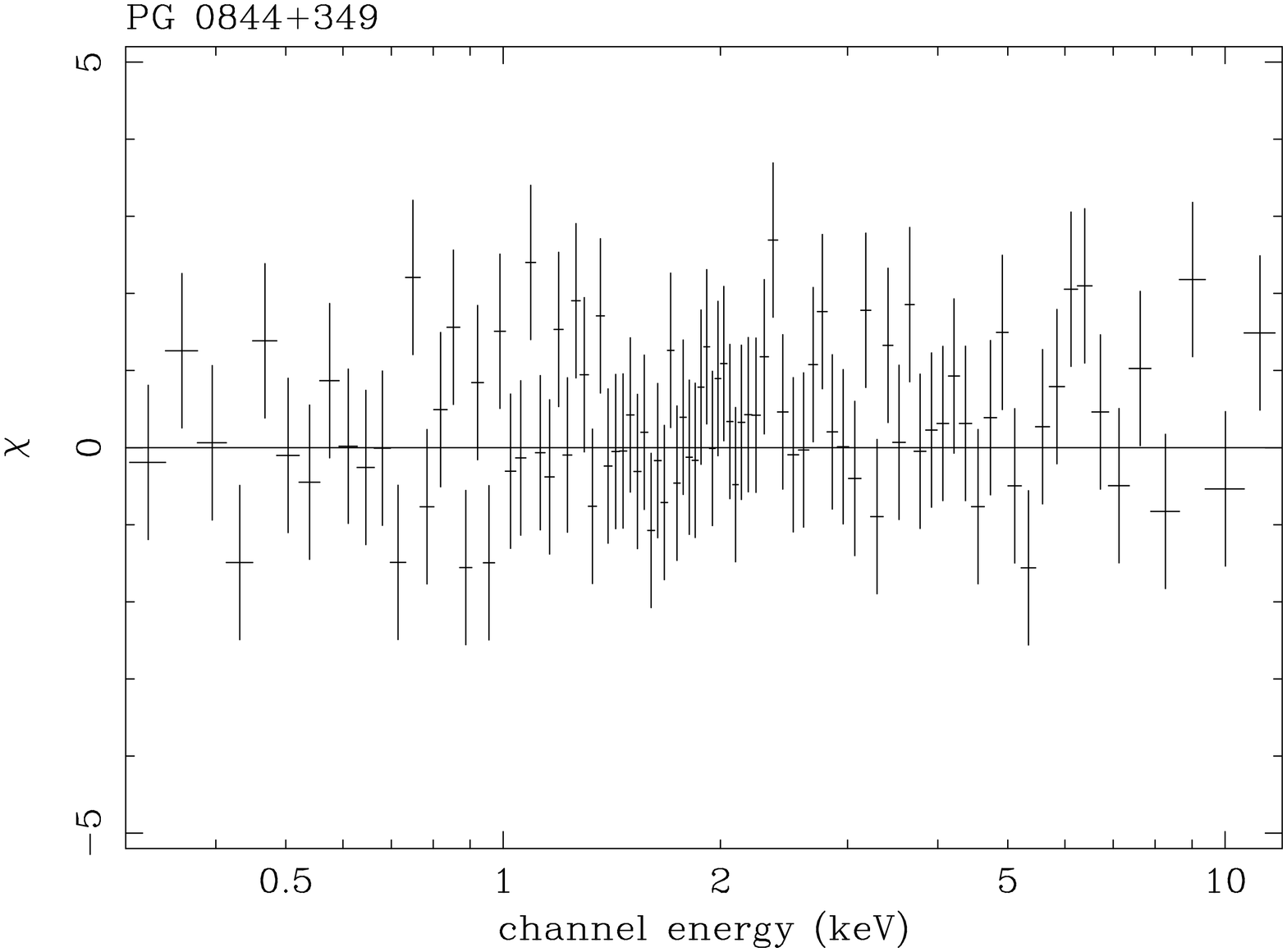}

\includegraphics[angle=270, width=0.4\textwidth]{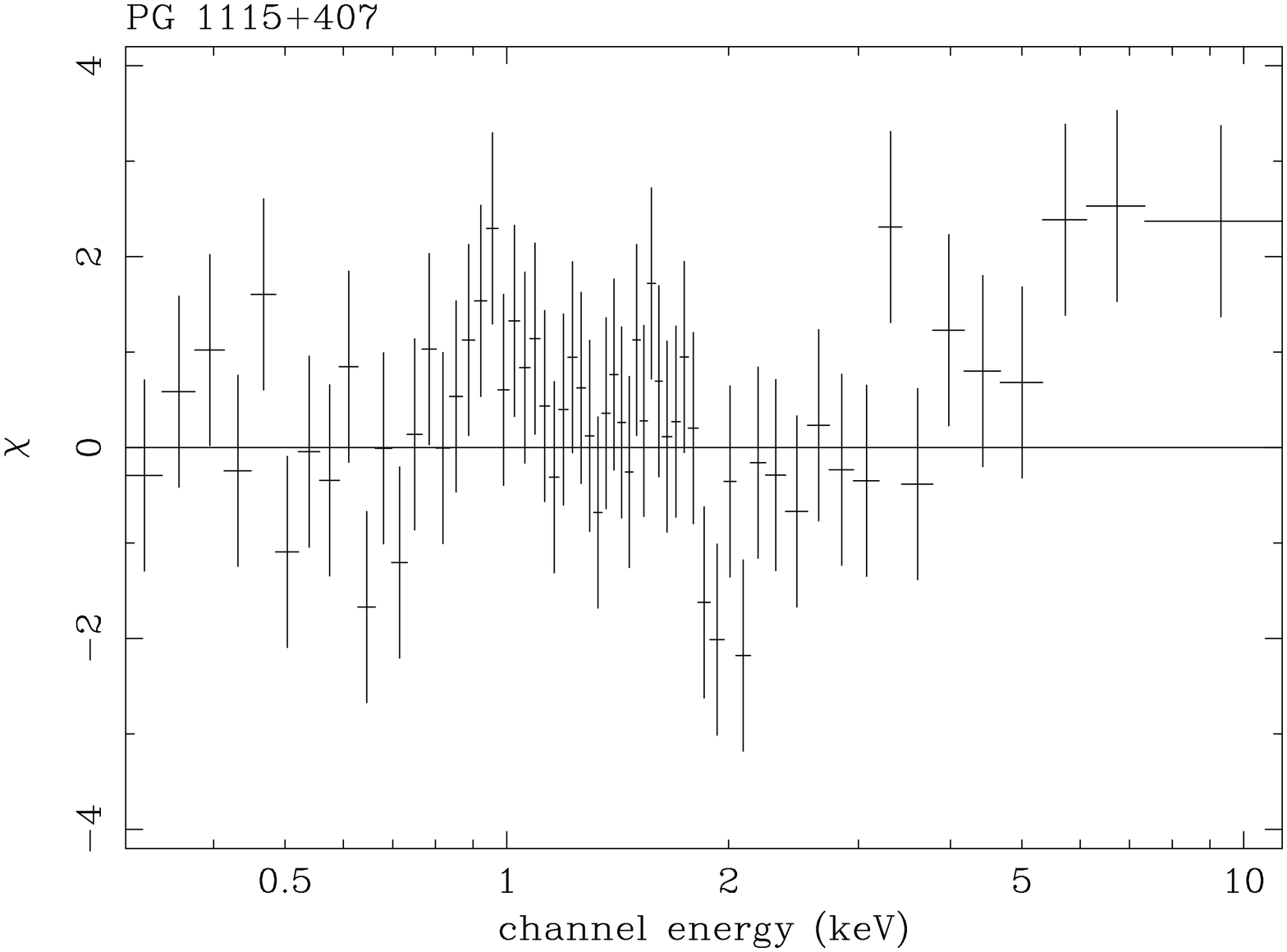}
\hspace{11mm}
\includegraphics[angle=270, width=0.4\textwidth]{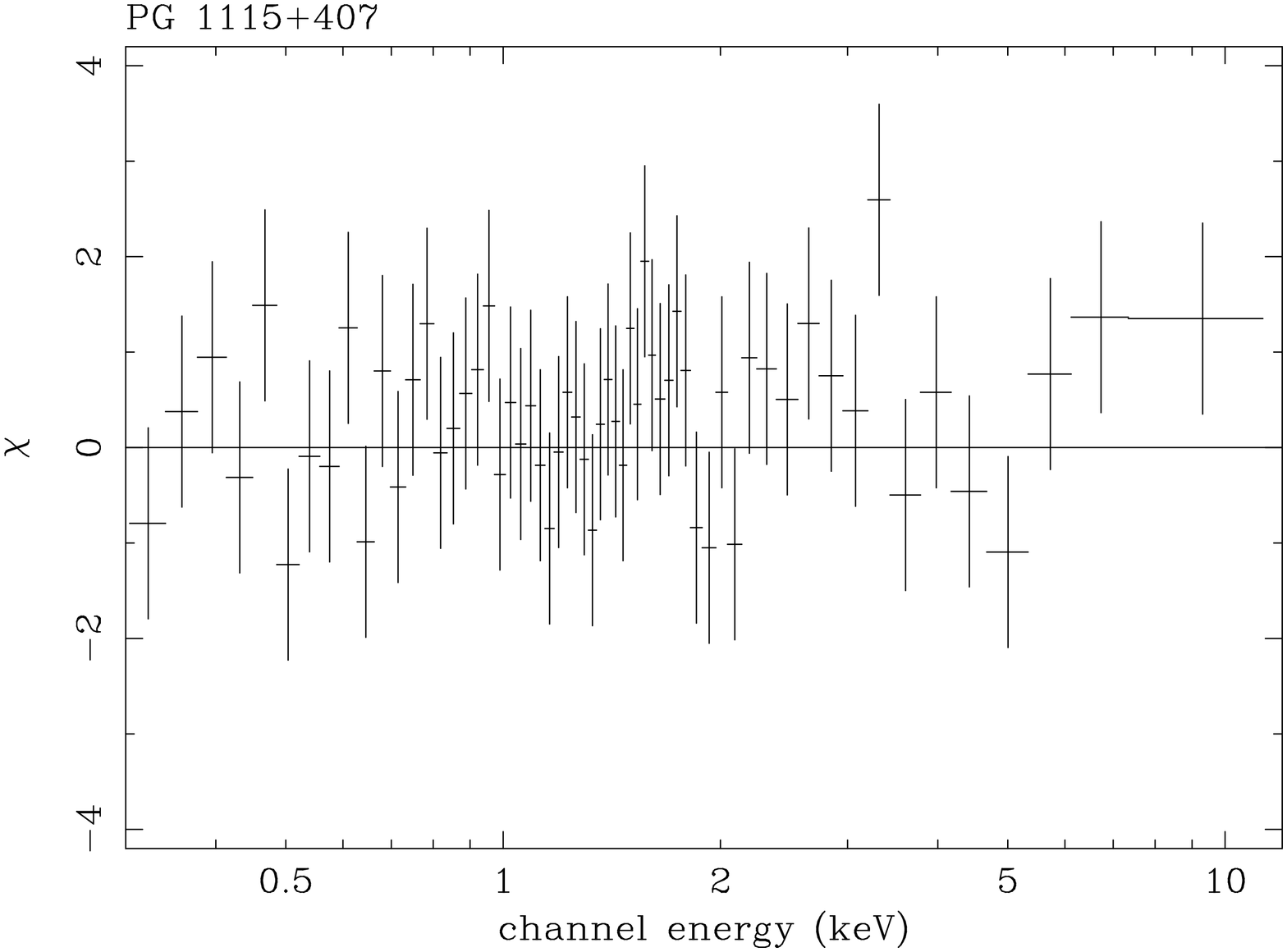}

\caption{The residuals to the simple model (left) and relativistically-blurred photoionized disc reflection model (right) fits to PG 0050+124, PG 0844+349 and PG 1115+407. The disc reflection model fit to PG 1115+407 does not include a power law component. The fit parameters are given in Tables \ref{phenom_fit_table}, \ref{disc_fit_table} and \ref{disc_edges_table}. The disc fits to these sources are an improvement on the simple fits, with $\chi^{2}_{\nu}$ (degrees of freedom) going from 1.270 (928) to 1.117 (924), 1.041 (631) to 0.987 (627) and 0.963 (416) to 0.913 (413) for PG 0050+124, PG 0844+349 and PG 1115+407, respectively.}
\label{residuals_figure}
\end{center}
\end{figure*}

\begin{figure*}
\begin{center}
\includegraphics[angle=270, width=0.4\textwidth]{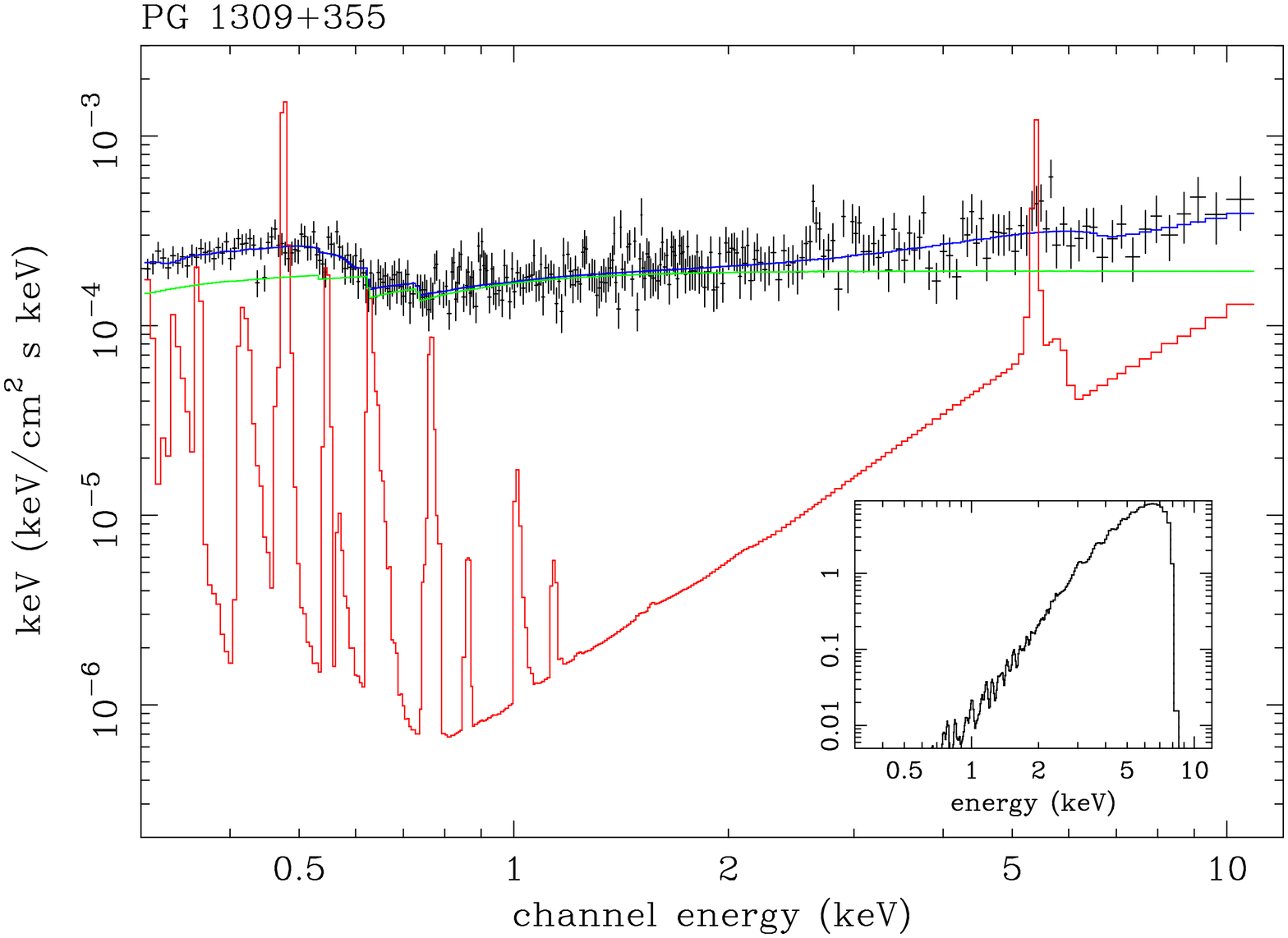}
\hspace{11mm}
\includegraphics[angle=270, width=0.4\textwidth]{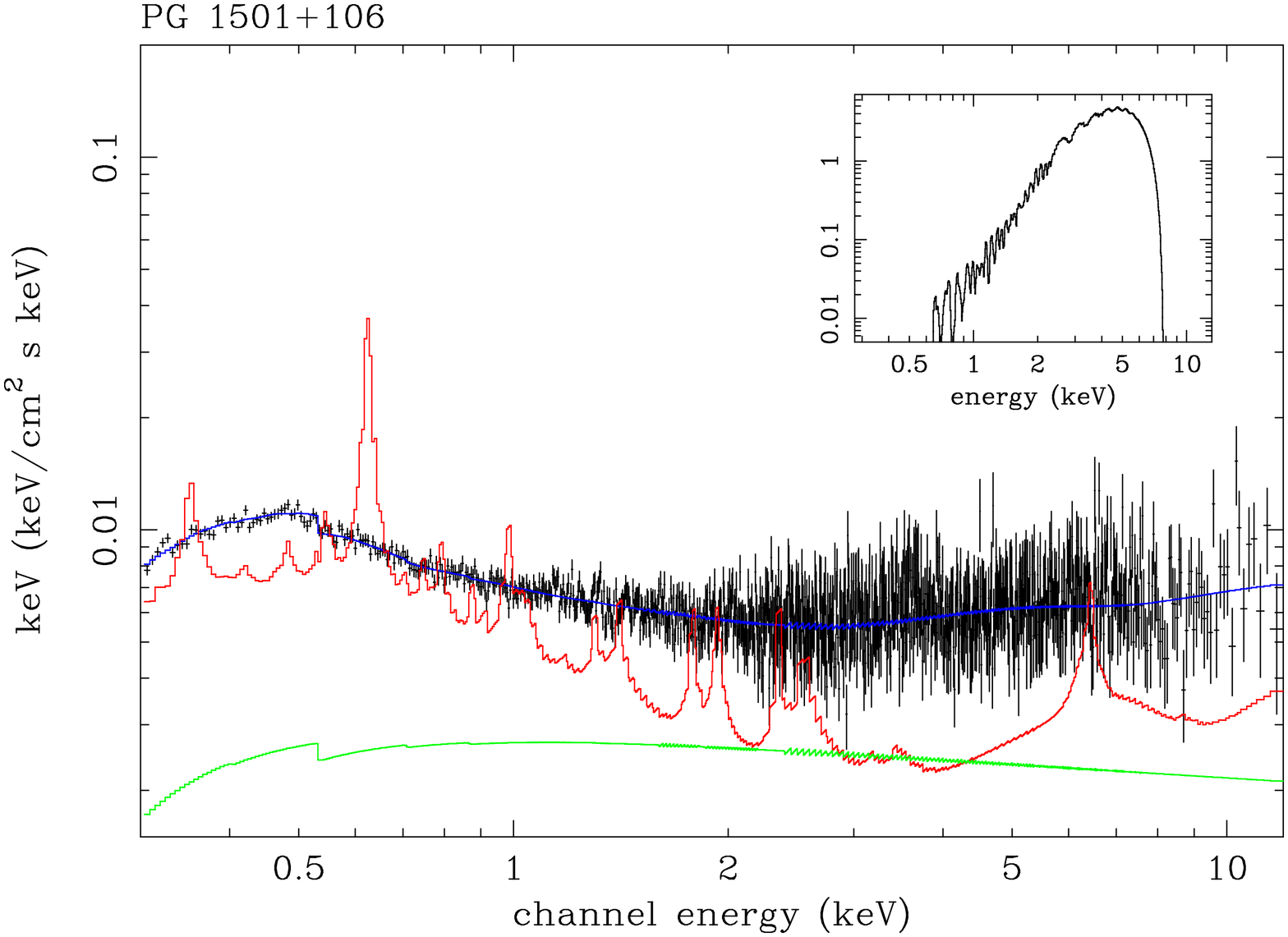}
\includegraphics[angle=270, width=0.4\textwidth]{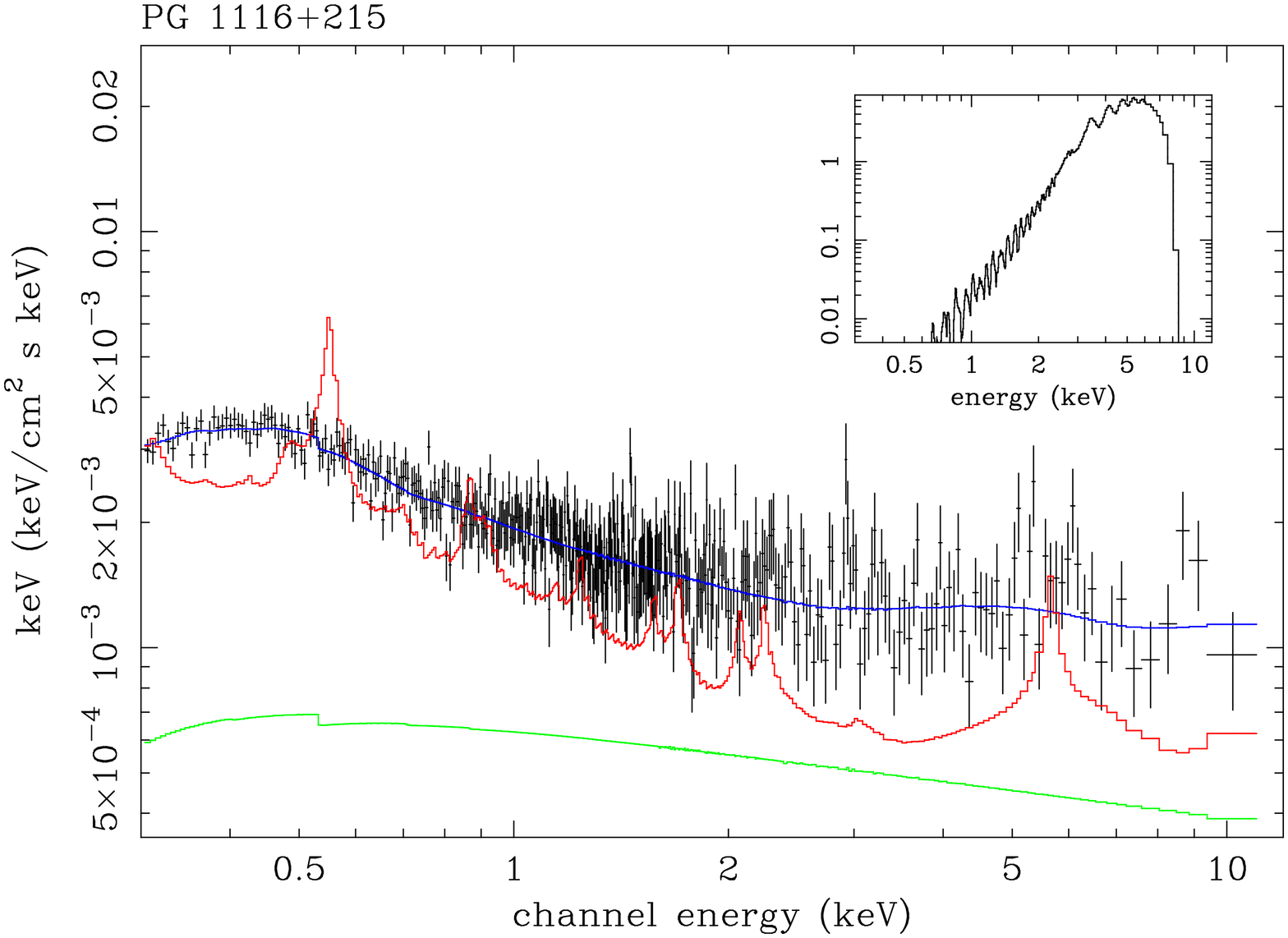}
\hspace{11mm}
\includegraphics[angle=270, width=0.4\textwidth]{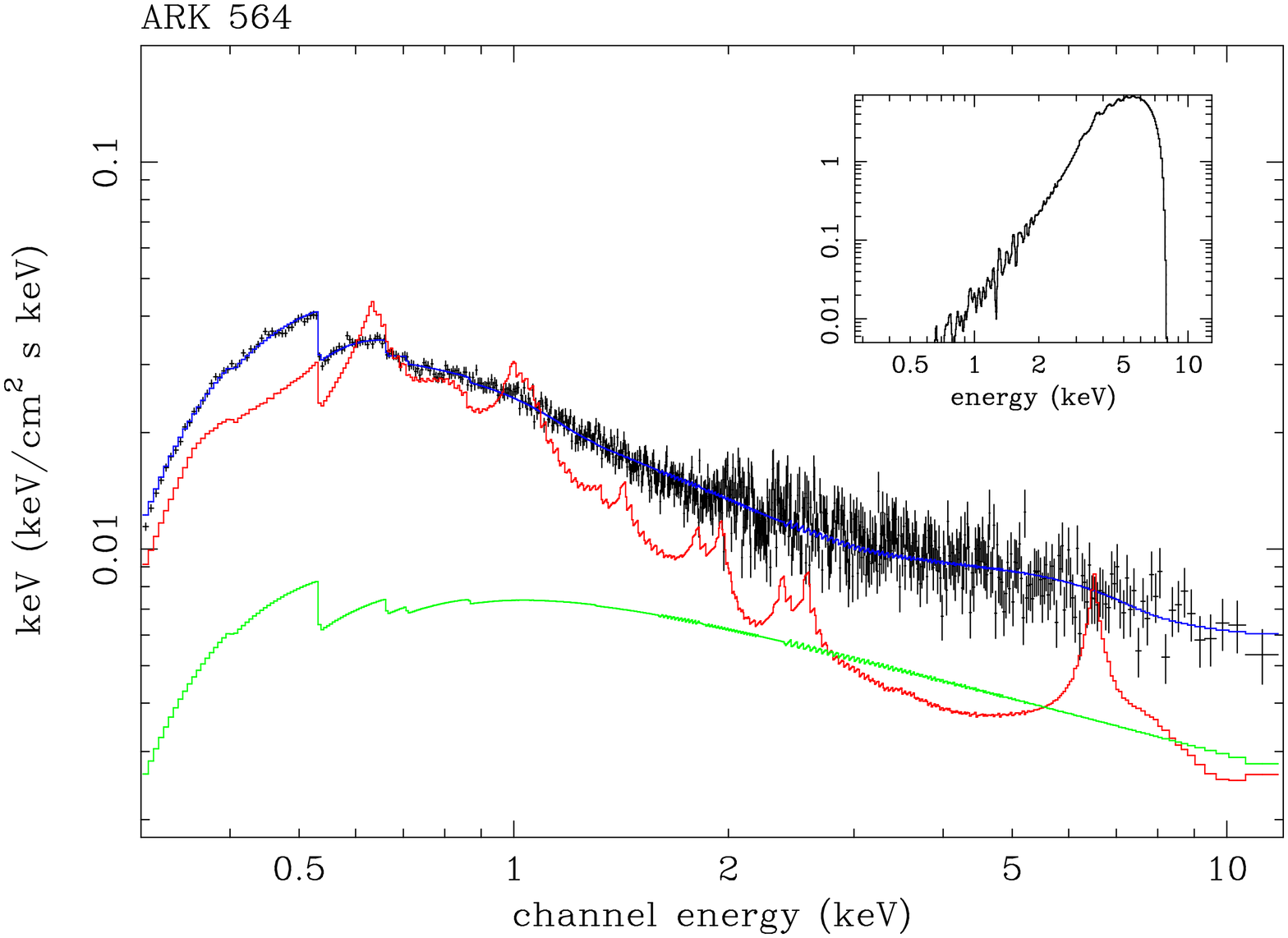}
\caption{The relativistically-blurred ionized disc reflection model plotted in $\nu F_{\nu}$, i.e. showing the flux emitted as a function of energy. The black crosses are the data points, the blue line is the fit to the data (from Table \ref{disc_fit_table}, with the absorption edges for PG 1309+355 (0.74~keV) and ARK 564 (0.68~keV) from Table \ref{disc_edges_table}), the green line is the power law component of the fit, the red line is the reflection component of the fit, and the insert shows the Laor profile (also plotted as  $\nu F_{\nu}$) that the two components are blurred with to produce the model. The figure illustrates the effect of ionization parameter ($\xi$) on the model, the model shown for PG 1309+355 has $\xi$ = 3, PG 1501+106 has $\xi$ = 510, PG 1116+215 has $\xi$ = 1270 and ARK 564 has $\xi$ = 3120. The lines are clearly more distinct in the reflection component at lower ionization parameters, and the Compton reflection continuum increases as ionization parameter increases. An iron line between 6.4 and 6.7~keV is clear in the reflection component for all values of the ionization parameter, but can only be seen in the source spectra as a slight enhancement to the continuum as it is blurred by the wide Laor lines (see Gallo et al. 2004b for an investigation of this feature in MRK 0586).}
\label{nufnu_figure}
\end{center}
\end{figure*}

We first fit the spectra with a power law and redshifted (local to the source) black body, modified by absorption from the cold gas in our Galaxy, with the amount of cold absorption fixed at the value given in Table \ref{sources_table}. We then allow for variable amounts of redshifted cold absorption (i.e. due to cold gas in the source galaxy) and warm absorption from ionized gas in the line of sight, and finally narrow iron emission due to e.g. reflection from a cold torus\footnote{In \textsc{xspec} terminology our simple model is \texttt{phabs*zphabs*zedge*zedge*(powerlaw+zbbody+zgauss)}}. We then fit the same data with the relativistically-blurred photoionized disc reflection model\footnote{Expressed in xspec commands our relativistically-blurred ionized disc reflection model is \texttt{phabs*zphabs*zedge*zedge*(kdblur(powerlaw+reflion)+zgauss)}, where \texttt{kdblur} is the relativistic convolution and \texttt{reflion} is a table model of the ionized reflection. Note that \texttt{reflion} includes redshift as a parameter.}, which we likewise check for the presence of redshifted cold and warm absorption, and narrow iron emission. We also investigate the possibility that the illuminating power law component is absent. Where adding a component produces an improvement in $\chi^{2}$ of $<$ 2.7 per lost degree of freedom we do not include that component in our final fit.\\
Specifically, to check for warm absorption we use two redshifted absorption edges whose energies are initially fixed at 0.74 and 0.87~keV to model the \textsc{O\thinspace VII} and \textsc{O\thinspace VIII} K absorption edges, as oxygen is the most abundant element in the soft X-ray band. We also allow the edge energy to vary to model in- or outflows, rejecting any fits outside 0.45 -- 1.1~keV as being unlikely to originate from \textsc{O\thinspace VII} or \textsc{O\thinspace VIII}. Further to this, given the recent controversy over whether AGN absorption edges are local to the sources or due to absorption from gas near our Galaxy (e.g. McKernan, Yaqoob \& Reynolds 2004) we also test the possibility of Galactic oxygen absorption with non-redshifted edges at 0.74 and 0.87~keV (we do not allow these energies to vary as relativistic outflows from our Galaxy are unlikely). We check for narrow iron emission using a redshifted Gaussian ($A(E) \propto e^{-(E-E_{0})^{2}/{2\sigma^{2}}}$) with line width $\sigma$ fixed at 5~eV (so the width is determined by the instrument resolution since the instrumental width is $\gg$ the intrinsic line width) and energy $E_{0}$ initially fixed at 6.4~keV (the energy at which neutral iron fluoresces most strongly), then allowing the energy to vary to account for the possibility of in- or outflows and ionized iron fluorescence. Since we are fitting the entire (0.3 -- 12.0~keV) spectrum whilst we test for the narrow iron line we only report lines for which \textsc{xspec} calculates a non-zero minimum equivalent width, this avoids the possibility that a detected line is due to a curvature in the continuum which our model has not addressed. Our technique is a conservative one and is likely to miss marginal lines, we only include lines which affect the overall goodness of fit. The comparison in goodness of fit between the two models will not be affected by any features that both models do not reproduce. A survey designed to detect fainter features (using the F-test), and including many of the same sources is available in Porquet et al. (2004).\\
We present the best black body and relativistically-blurred photoionized disc reflection model fits in Tables \ref{phenom_fit_table} and \ref{disc_fit_table}, respectively. A plot of our measured soft excess temperature is shown in Fig. \ref{soft_temp_figure}, and several example comparisons between the two different model fits are plotted in Figs. \ref{residuals_figure}, \ref{ngc4051_spectrum} and \ref{1211}. Example $\nu F_{\nu}$ plots are shown in Fig. \ref{nufnu_figure}.\\
We find cold absorption in only one source, PG 0050+124, and clear narrow iron lines in two, NGC 4051 and MRK 766. The same sources show extra cold absorption local to the source and narrow iron lines with either model, with the exception of PG 1211+143 which only appears to have an iron line when the simple model is used. PG 1211+143 shows spectral complexity in the 4 -- 8~keV region, see Section \ref{notes_section} and Fig. \ref{1211}. We find redshifted warm absorption edges in 18 sources using the simple model, compared to 7 sources using the disc reflection model. We find non-redshifted warm absorption in only one source using the simple model, and three sources using the disc reflection model. We do not detect a power law component in 11 of the 34 sources when using the disc reflection model.\\
Note that the disc reflection model has some limits built in. The Laor profile has an inner disc radius that cannot be $<$ 1.235 gravitational radii, for physical reasons discussed in Section \ref{discussion_section}. The profile also has a disc emissivity index which cannot be $>$ 10, this is quite reasonable as an emissivity profile going as the tenth power of radius (n.b. where $r$ is radius and $A$ is area $r^{-10} dA \propto r^{-9} dr$ since $dA/dr = 2\pi r$ for a disc) emits 90 per cent of its radiation between 1.235 (the inner edge) and 1.65 gravitational radii (this may not equal the fraction that reaches the observer due to various relativistic effects, see Section \ref{discussion_section}). The reflection model is a grid model which is restricted in parameter space to increase the speed with which it can be computed. The limits in our version of the model which the sources encounter in fitting are the constraint that iron abundance (Fe) must be $<$ 10 times solar and the ionization parameter ($\xi$) is restricted to 1 $< \xi <$ 10000~erg cm s$^{-1}$.\\
It is also possible to create a disc reflection model based on a non-rotating black hole, by using the diskline (Fabian et al. 1989) profile instead of the Laor profile for the blurring. This model is less successful at fitting the data; whilst it matches the general shape of the spectra it is less smooth than the Laor-blurred model and leaves residuals. We also fit using this diskline model, for all the sources the fit is worse than the Laor based model and is completely inadequate ($\chi^{2}_{\nu} > 2$) in several cases. We report sources where the fit is not significantly worse in Table \ref{nonrotating_fit_table}.

\subsection{Reflection fraction and flux fraction}\label{fluxfrac_section}
\begin{figure}
\begin{center}
\includegraphics[angle=0, width=0.45\textwidth]{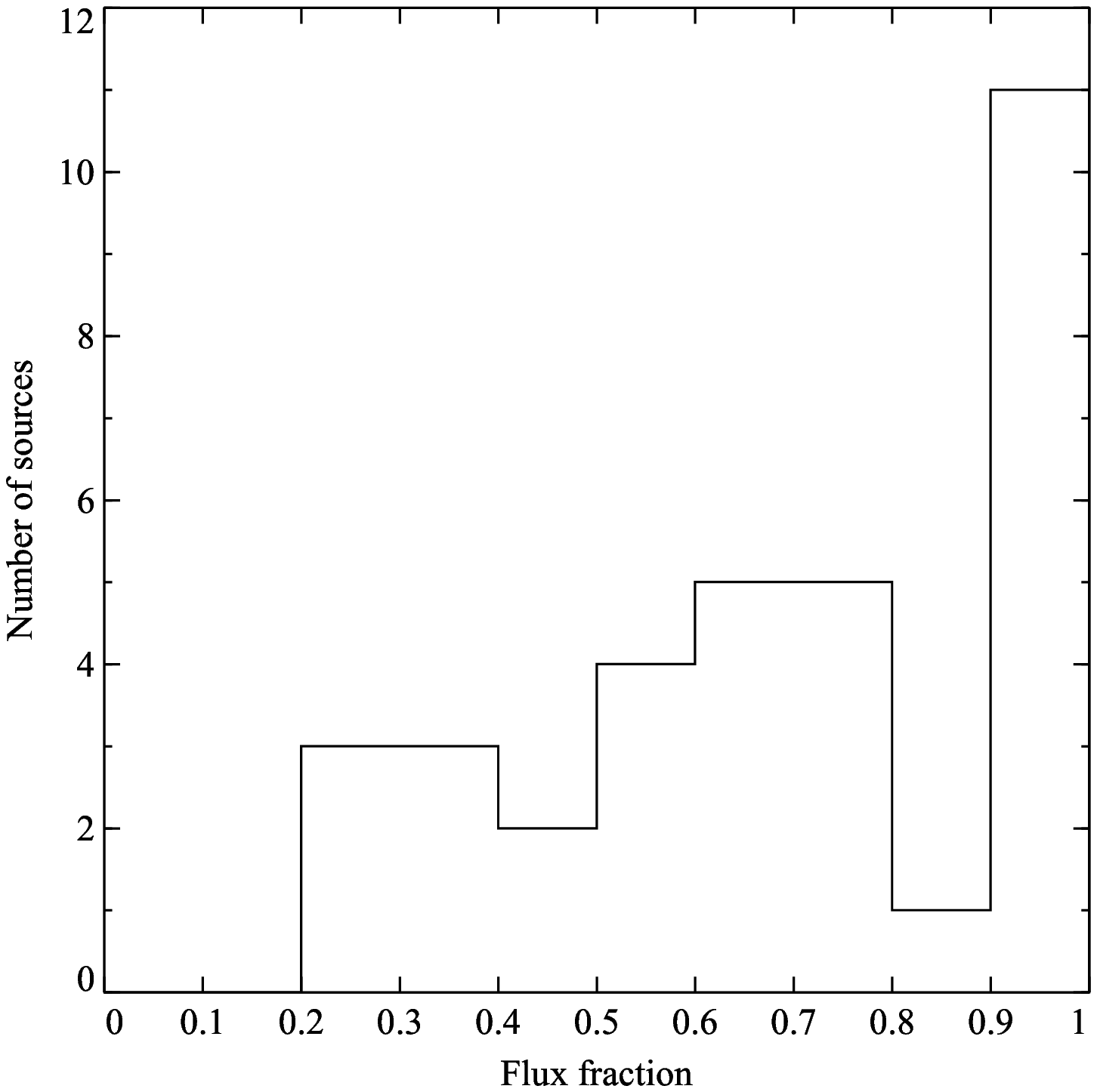}
\caption{A histogram showing the measured flux fractions of the sources. All of the sources have a reflection component, and many sources are dominated by reflection. The deficit of sources in the range 0.8 -- 0.9 may be partly due to low data quality, see Section \ref{results}.}
\label{flux_frac_figure}
\end{center}
\end{figure}

Conventionally, the reflection fraction is the ratio of reflected emission to power law emission. It is based on the assumption of an isotropic source of power law emission, a certain fraction of which is reflected from gas. The reflection fraction is then trivially related to the solid angle at the source of power-law emission which is covered by reflecting gas. An infinite plane of gas illuminated by an isotropic source would have a reflection fraction of 0.5. A source with no reflection has a reflection fraction of zero and a source with no power law emission has an infinite reflection fraction. Reflection fractions above one show that the AGN geometry is more complicated than a source and some clouds of gas, implying e.g. anisotropic emission or obscuration of the power law source.\\
In the vicinity of a black hole, many of the fundamental assumptions used to calculate a reflection fraction and extract physical meaning from it are known to be violated. The matter in the disc is moving at relativistic speeds, so reflected radiation from the disc tends to be Doppler beamed in the plane of the disc, violating isotropy. In a non-relativistic situation all the energy that is incident on the disc is reprocessed and reemitted, but with Doppler boosting the reemitted radiation as detected by an observer may pick up more energy from the motions of the disc. Gravitational redshifting likewise makes comparing emitted and reflected energy non-trivial. Finally, space is curved, and light no longer propagates in straight lines -- even if the source of power law emission is isotropic and not moving at relativistic speeds (i.e. with no beaming), the light bending effects will ensure that more radiation impacts on the disc than is detected by an observer, again violating isotropy (e.g. Miniutti \& Fabian 2004). All of these effects may either increase or decrease the observed reflection fraction, and complicate the situation to such a degree that the conventional reflection fraction becomes physically meaningless. We therefore report only the purely observational quantity we call the flux fraction, which we define as the measured fraction of reflected flux over the total observed flux.\\
We measure the flux fraction for each of the fits listed in Table \ref{disc_fit_table}. We use the `flux' command from \textsc{xspec} to calculate the total flux of the model over the range 0.3 -- 12.0~keV, setting all warm and cold absorption and narrow Gaussian emission to zero so as to measure the flux emitted by the inner portion of the AGN, then set the power law component to zero and repeat the measurement to find the amount of reflected radiation. The flux fraction is the ratio of reflected emission to total emission, so a source with no observed reflection has a flux fraction of zero and a source with no observed power law emission has a flux fraction of one. The previously mentioned infinite plane of gas illuminated by an isotropic source, with no relativistic effects, would have a flux fraction of approximately 0.5 (not exactly 0.5 as reflection can shift photons outside the observed band). The flux fraction is dependent on the energy range it is measured over, so we use the same limits as our spectral fitting. The flux fractions are shown in Figure \ref{flux_frac_figure}.\\
We find that no source in our sample has a flux fraction below 0.25, that the flux fraction is roughly evenly distributed between 0.25 and 0.8, and that $\sim$ one third of our sources have no power law component. This extreme reflection may be due to several effects, discussed in Section \ref{model_section}. We only measure one flux fraction above 0.8 and below 1.0 (with 11 sources having a flux fraction of 1.0), this may be partly due to our fitting procedure; adding a power law component increases the number of degrees of freedom by one, so it must improve $\chi^{2}$ by $\geq$ 2.7 for us to judge it significant and include it. Small power law components therefore tend not to be included, e.g. in our fit to PHL 1092 (a low quality observation where our best fit does not include a power law component) adding a power law component such that the flux fraction is 0.98 worsens the fit by a $\Delta \chi^{2}$ of 2, so a small power law component does not affect the fit enough to be significant. However, adding a power law component such that the flux fraction is 0.9 makes the fit worse by $\Delta \chi^{2}$ of 11, enough to be clearly ruled out as a worse fit. The deficit in measured flux fractions between 0.8 and 0.9 is therefore not due to our fitting methods, but it may still be an artefact of our small sample size. It is also possible that there are two populations of AGN: those with flux fractions of 0.2 -- 0.8 and those totally dominated by reflection.\\

\begin{table*}
\caption{Simple fits to the sources. $\Gamma$ is power law spectral index, $kT$ is black body temperature, Edge Energy and $\tau$ are the energy and optical depth of the two absorption edges. Energies given without associated errors are fixed at 0.74 or 0.87~keV, the energy of the \textsc{O\thinspace VII} and \textsc{O\thinspace VIII} K absorption edges. Blank spaces indicate no feature is present.}
\label{phenom_fit_table}
\begin{tabular}{llllllll}
\hline
Source & $\Gamma$ & $kT$ (eV) & Edge Energy (keV)& $\tau$ & Edge Energy (keV)& $\tau$ &  $\chi^{2}_\nu $ (d.o.f.)\\
\hline

PG 0003+199 & $2.182^{+0.007}_{-0.008} $ & $ 132^{+1}_{-1} $ & $0.640^{+0.008}_{-0.010} $ & $ 0.24^{+0.02}_{-0.02} $ & & & $1.065   (970)$\\

PG 0050+124${}^{a}$ & $	2.331^{+0.009}_{-0.007} $ & $130^{+30}_{-30} $ & & & & & $1.270  (928) $\\ 

PG 0157+001 & $2.12^{+0.13}_{-0.05} $ & $130^{+10}_{-20} $ & & & & & $0.870 (237)$\\

PG 0844+349 & $ 2.19^{+0.03}_{-0.03} $ & $116^{+2}_{-1} $ & & & & & $1.041  (631)$ \\

PG 0947+396 & $1.93^{+0.04}_{-0.05} $ & $170^{+10}_{-10} $ &	0.74 & $0.28^{+0.09}_{-0.08} $ & 0.87 & $ 0.12^{+0.11}_{-0.09} $ & $ 0.943 (479)$ \\

PG 0953+414 & $2.12^{+0.04}_{-0.04} $ & $151^{+7}_{-7} $ & 0.74 & $0.31^{+0.09}_{-0.08} $ & & & $1.096  (515)$ \\

PG 1048+342 & $1.90^{+0.05}_{-0.05} $ & $134^{+6}_{-7} $ & & & & & $1.075 (538)$ \\

PG 1115+407 & $2.44^{+0.05}_{-0.05} $ & $111^{+5}_{-5} $ & & & & & $ 0.963 (416)$ \\

PG 1116+215 & $ 2.28^{+0.05}_{-0.04} $ & $112^{+5}_{-6} $ & & & & & $ 1.048  (401)$ \\

PG 1202+281 & $1.76^{+0.04}_{-0.04} $ & $160^{+9}_{-9} $ & 0.74 & $ 0.33^{+0.09}_{-0.08} $ &	 & & $1.036  (554)$ \\

PG 1211+143${}^{b}$ & $ 1.789^{+0.009}_{-0.009} $ & $ 116.4^{+0.6}_{-1.0} $ & $ 0.754^{+0.007}_{-0.006} $ & $ 0.40^{+0.03}_{-0.02} $ & $ 0.94^{+0.01}_{-0.01} $ & $ 0.34^{+0.03}_{-0.02} $ & $ 1.189 (935) $ \\

PG 1244+026 & $2.63^{+0.05}_{-0.06} $ & $154^{+7}_{-3} $ & & & & & $ 1.137 (295)$ \\

PG 1307+085 & $1.51^{+0.07}_{-0.05} $ & $ 130^{+20}_{-20} $ & 0.74 & $ 0.6^{+0.2}_{-0.2} $ & & & $ 0.893 (269) $\\

PG 1309+355 & $1.71^{+0.05}_{-0.03} $ & $110^{+20}_{-10} $ & 0.74 & $0.4^{+0.2}_{-0.2} $ & & & $0.916 (298)$ \\

PG 1322+659 & $2.33^{+0.06}_{-0.06}$ & $113^{+4}_{-4} $ & & & & & $ 1.041  (335)$ \\

PG 1352+183 & $ 2.21^{+0.05}_{-0.05} $ & $111^{+7}_{-7} $ & & & & & $1.089  (373)$\\

PG 1402+261 & $ 2.35^{+0.05}_{-0.05} $ & $116^{+5}_{-5} $ & & & & & $1.220  (387)$\\

PG 1404+226 &  $ 1.66^{+0.32}_{-0.25} $ & $116^{+3}_{-3} $ & 0.87 & $0.4^{+0.1}_{-0.2} $ & & & $0.950 (192) $\\

PG 1427+480 & $2.00^{+0.04}_{-0.04} $ & $156^{+7}_{-6} $ & 0.74	& $0.23^{+0.08}_{-0.08} $ & & & $1.013  (527)$\\

PG 1440+356 & $2.51^{+0.08}_{-0.07} $ & $ 101^{+4}_{-4} $ & & & & & $1.057 (292)$ \\

PG 1444+407 & $2.35^{+0.08}_{-0.08} $ & $150^{+10}_{-10} $ &	0.74 & $0.3^{+0.1}_{-0.1} $ & & & $ 1.326 (291)$ \\

PG 1501+106 & $2.07^{+0.02}_{-0.02} $ & $ 102^{+2}_{-2} $ & & & & & $1.177 (935)$ \\

NGC 4051${}^{b}$ & $ 2.006^{+0.002}_{-0.004} $ & $ 109.9^{+0.2}_{-0.4} $ & $ 0.711^{+0.006}_{-0.005} $ & $ 0.120^{+0.008}_{-0.004} $ & & & $ 2.138 (1045) $ \\

IRAS 13349+2438 & $2.08^{+0.01}_{-0.01} $ & $97.2^{+0.8}_{-0.7} $ & 0.74 & $0.37^{+0.03}_{-0.03} $ & $1.03^{+0.03}_{-0.34} $ & $ 0.17^{+0.04}_{-0.03} $ & $ 1.122 (670)$ \\ 

ARK 564 & $2.60^{+0.03}_{-0.01} $ &	$133^{+2}_{-2} $ & $0.675^{+0.009}_{-0.006} $ & $ 0.24^{+0.02}_{-0.03} $ & & & $ 1.287 (670)$ \\

MRK 1044 & $2.20^{+0.03}_{-0.04} $ & $107^{+2}_{-2} $ & & & & & $ 1.104 (534)$ \\

E 1346+266 & $ 2.69^{+0.06}_{-0.11} $ & $164^{+7}_{-7} $ & & & & & $ 1.058 (233)$ \\

MRK 0359 & $ 1.92^{+0.04}_{-0.04} $ & $128^{+4}_{-4} $ & & & & & $ 1.176 (530)$ \\

PHL 1092 & $2.09^{+0.17}_{-0.08} $ & $ 93^{+2}_{-2} $ & & & & & $ 1.174 (183) $ \\

RE J1034+396 & $2.5^{+0.2}_{-0.1} $ & $144^{+3}_{-4} $ & $ 0.60^{+0.02}_{-0.02} $ & $ 0.45^{+0.08}_{-0.10} $ & $ 0.83^{+0.02}_{-0.03} $ & $0.45^{+0.09}_{-0.07}$ & $ 1.047 (281) $ \\

PKS 0558-504 & $ 2.32^{+0.01}_{-0.01} $ & $127^{+2}_{-2} $ & $ 0.60^{+0.02}_{-0.01} $ & $0.23^{+0.03}_{-0.03} $ & & & $ 1.287 (375) $ \\

MRK 0766${}^{b}$ &  $ 2.010^{+0.005}_{-0.005} $ & $ 103^{+1}_{-1} $ & $ 0.71^{+0.004}_{-0.005} $ & $ 0.39^{+0.02}_{-0.02} $ & & & $ 1.123 (956) $ \\

MRK 0586 & $2.40^{+0.03}_{-0.04} $ & $127^{+2}_{-3} $ &  0.74 & $0.13^{+0.06}_{-0.03} $ & & & $ 1.027 (526)$  \\

TON S180 & $2.52^{+0.09}_{-0.09} $ & $ 114^{+1}_{-1} $ & $0.460^{+0.006}_{-0.008} $ & $ 0.35^{+0.02}_{-0.02} $& & & $ 1.298 (780)$\\

\hline

\end{tabular}
\\
${}^{a}$ The fit includes redshifted cold absorption with an equivalent hydrogen column of $ 3.8^{+0.2}_{-0.1} \times 10^{20}$~cm$^{-2}$. ${}^{b}$ These fits include a narrow Gaussian line at 6.4~keV, equivalent width $80^{+40}_{-60}$~eV for PG1211+143, $109^{+18}_{-20}$~eV for NGC 4051 and $70^{+40}_{-30}$~eV for MRK 0766.
\end{table*}

\begin{table*}
\caption{Relativistically-blurred photoionized disc fits to the sources. Em. Ind. is the index of the emissivity of the accretion disc (assumed to be a power law), $R_{in}$ is the disc's inner radius (in gravitation radii), $i$ is the inclination of the disc to the line of sight (in degrees), Fe is the iron abundance of the disc (relative to solar), $\xi$ (in erg cm s$^{-1}$) is the ionization parameter of the gas ($\xi = 4 \pi F/n_{H}$, where $F$ is the illuminating energy flux and $n_{H}$ is the hydrogen number density in the illuminated layer). Flux Frac. is the flux fraction, see Section \ref{results}}\label{disc_fit_table}
\begin{tabular}{lllllllllll}
\hline
  Source & Em. Ind. &  $R_{in}$ & $i$ &  Fe & $\Gamma$ &  $\xi$ & $\chi^{2}_{\nu}$ (d.o.f.)  & Flux Frac. & \\
\hline
 PG 0003+199  & $ 5.2	^{+0.6}_{-0.2}	 $ & $ 1.6	^{+0.4}_{-0.1}	 $ & $ 58	^{+1}_{-4}	 $ & $ 0.71	^{+0.04}_{	 -0.06}	 $ & $ 2.189	^{+0.005}_{	-0.015}	 $ & $ 1020 ^{+40}_{  -80}  $ & $ 0.939 (968) $ & $ 0.71 \pm 0.04 $  \\

 PG 0050+124${}^{a}$ & $ 4.6^{+0.6}_{-0.5} $ & $ 4.0^{+0.5}_{-0.4} $ & $ 22^{+2}_{-2} $ & $ 0.46^{+0.05}_{-0.06} $ & $ 2.36^{+0.02}_{-0.01} $ & $ 1150^{+120}_{-90} $ & $ 1.117 (924) $ & $ 0.34 \pm 0.07 $ \\

PG 0157+001${}^{b}$ & $ 8.5^{+1.5}_{-4.7} $ & $ 1.2^{+0.3}_{-0.0} $ & $ 73^{+7}_{-15} $ & $ 0.5^{+0.4}_{-0.2} $ & $ 2.05^{+0.09}_{-0.07} $ & $ 1250^{+870}_{-290} $ & $ 0.861 (234) $ & 1 \\

PG 0844+349 & $ 10.0^{+0.0}_{-1.8} $ & $ 1.6^{+1.1}_{-0.1} $ & $ 59^{+3}_{-11} $ & $ 0.72^{+0.09}_{-0.14} $ & $ 2.19^{+0.04}_{-0.02} $ & $ 520^{+190}_{-100} $ & $ 0.987 (627) $ & $ 0.53 \pm 0.15$  \\

 PG 0947+396  & $ 10.0^{+0.0}_{-2.8} $ & $ 1.4 ^{+0.5}_{-0.2} $ & $ 71 ^{+8}_{-20} $ & $ 0.5 ^{+0.1}_{-0.1} $ & $ 1.90^{+0.06}_{-0.06} $ & $ 710 ^{+360}_{-210} $ & $ 0.893 (477) $ & $ 0.74 \pm 0.24 $    \\

PG 0953+414 & $ 6.2^{+1.0}_{-2.3} $ & $ 1.3^{+0.3}_{-0.1} $ & $ 62^{+11}_{-5} $ & $ 0.8^{+1.0}_{-0.4} $ & $ 2.35^{+0.03}_{-0.03} $ & $ 100^{+10}_{-40} $ & $ 1.055 (512) $ & $ 0.28 \pm 0.16 $  \\

 PG 1048+342 & $ 7.7 ^{+0.8}_{-1.8} $ & $ 1.2 ^{+0.4}_{-0.0} $ & $ 69 ^{+5}_{-6} $ & $ 0.7 ^{+0.2}_{-0.1} $ & $1.86^{+0.02}_{-0.06}$ & $ 600 ^{+250}_{-100} $ & $ 1.043 (534) $ & $ 0.68 \pm 0.16 $  \\

PG 1115+407${}^{b}$ & $ 8.9^{+1.1}_{-3.5} $ & $ 1.2^{+0.2}_{-0.0} $ & $ 63^{+6}_{-15} $ & $ 0.4^{+0.1}_{-0.1} $ & $ 2.30^{+0.03}_{-0.03} $ & $ 2320^{+840}_{-500} $ & $ 0.913 (413) $ & 1 \\

 PG 1116+215  & $ 3.6	^{+2.0}_{-0.5} $ & $ 1.3	^{+2.0}_{-0.0} $ & $ 67	^{+5}_{-2} $ & $ 0.5	^{+2.1}_{-0.2} $ & $ 2.41^{+0.08}_{-0.03} $ & $ 1270 ^{+300}_{-460} $ & $ 1.007 (397) $ & $ 0.70 \pm 0.35 $   \\

 PG 1202+281 & $ 9.6 ^{+0.4}_{-4.9} $ & $ 1.2 ^{+0.3}_{-0.0} $ & $ 71 ^{+4}_{-16} $ & $ 1.3 ^{+1.2}_{-0.8} $ & $ 2.03 ^{+0.02}_{-0.04} $ & $ 30 ^{+10}_{-15} $ & $1.009 (553) $ & $ 0.26 \pm 0.03 $  \\

 PG 1211+143${}^{c}$  &  $ 7.8^{+0.6}_{-1.7} $ & $ 1.24^{+0.08}_{-0.00} $ & $ 40^{+2}_{-10}  $ & $ 10.0^{+0.0}_{-0.5}  $ & $ 1.83^{+0.02}_{-0.03}  $ & $ 370^{+10}_{-10}  $ & $ 1.274 (936)  $ & $ 0.70 \pm 0.09 $   \\ 

PG 1244+026${}^{b,d}$ & $5.6^{+4.4}_{-2.7} $ & $ 2.3^{+5.8}_{-1.1} $ & $ 45^{+23}_{-45} $ & $ 1.4^{+0.4}_{-0.3} $ & $ 2.42^{+0.04}_{-0.03}$ & $ 10000^{+0.0}_{-2800} $ & $ 1.102 (292) $ & 1  \\

 PG 1307+085${}^{e}$ & $8.4^{+1.6}_{-4.8} $ & $1.3^{+0.6}_{-0.1} $ & $ 64^{+6}_{-26} $ & $ 2.6^{+4.4}_{-1.9} $ & $ 1.83^{+0.01}_{-0.07} $ & $ 40 ^{+130}_{-10} $ & $ 0.871 (265) $ & $ 0.40 \pm 0.25 $ \\

 PG 1309+355${}^{e}$  & $7.6^{+2.4}_{-4.2} $ & $ 1.3^{+1.7}_{-0.1} $ & $ 51 ^{+16}_{-25} $ & $ 1.0^{+9.0}_{-0.7} $ & $1.8^{+0.1}_{-0.1} $ & $3^{+57}_{-2} $ & $ 0.927 (293) $ & $ 0.25 \pm 0.42 $  \\

 PG 1322+659 & $4.5^{+2.3}_{-1.0}$  & 	$1.4^{+1.1}_{-0.2} $ & $34   ^{+27}_{-34} $& $0.9  ^{+0.3}_{-0.2} $& $2.33   ^{+0.05}_{-0.06}$ & $1400  ^{+550}_{-300} $ & $1.009 (331)$ & $ 0.67 \pm 0.24 $  \\

PG 1352+183${}^{b}$ & $ 4.6^{+2.4}_{-1.4} $ & $ 1.6^{+0.9}_{-0.3} $ & $ 35^{+17}_{-35} $ & $ 0.4^{+0.1}_{-0.1} $ & $ 2.13^{+0.03}_{-0.03} $ & $ 1450^{+300}_{-210} $ & $ 1.002 (370) $ & 1  \\

PG 1402+261${}^{b}$ & $ 9.5^{+0.5}_{-3.0} $ & $ 1.2^{+0.1}_{-0.0} $ & $ 66^{+3}_{-11} $ & $ 0.5^{+0.1}_{-0.1} $ & $ 2.20^{+0.04}_{-0.01} $ & $ 1340^{+80}_{-120} $ & $ 1.129 (384) $ & 1 \\

PG 1404+226${}^{b}$ & $ 6.4^{+1.8}_{-1.8} $ & $ 1.2^{+0.4}_{-0.0}  $ & $ 46^{+13}_{-18} $ & $ 4.4^{+1.8}_{-1.1} $ & $ 2.3^{+0.2}_{-0.1} $ & $ 810^{+270}_{-380} $ & $0.893 (190)$ & 1 \\

 PG 1427+480  & $   4.8  ^{ +1.0}_{  -0.8} $ & $ 1.31 ^{  +0.35}_{  -0.08} $ & $ 72   ^{+2}_{   -22} $ & $ 0.7  ^{+0.2}_{ -0.1} $ & $ 2.01   ^{+0.07}_{  -0.05} $ & $ 800   ^{+280}_{  -150} $ & $ 0.985 (524) $ & $ 0.70 \pm 0.25 $  \\


PG 1440+356${}^{b}$ & $ 9.8^{+0.2}_{-3.1} $ & $ 1.2^{+0.1}_{-0.0} $ & $ 59^{+3}_{-24} $ & $ 0.5^{+0.2}_{-0.1} $ & $ 2.35^{+0.02}_{-0.03} $ & $ 1770^{+400}_{-360} $& $ 0.989 (289) $ & 1 \\

PG 1444+407${}^{b}$ & $ 4.0^{+4.8}_{-1.0} $ & $ 1.4^{+1.7}_{-0.2} $ & $40^{+19}_{-16} $ & $ 0.6^{+0.2}_{-0.1} $ & $ 2.28^{+0.05}_{-0.09} $ & $ 2310^{+960}_{-960} $ & $ 1.297 (289) $ & 1 \\

 PG 1501+106 & $ 9.7^{+0.3}_{-2.3} $ & $1.2^{+0.2}_{-0.0} $ & $58^{+2}_{-13} $ & $ 0.40^{+0.03}_{-0.03} $ & $2.12^{+0.02}_{-0.04} $ & $ 510^{+160}_{-110} $ & $ 0.981 (931) $& $ 0.67 \pm 0.17 $ \\

 NGC 4051${}^{e}$   & $ 5.77 ^{+0.02}_{-0.51} $ & $ 1.24^{+0.04}_{-0.00} $ & $ 40.0^{+0.4}_{-0.2} $ & $ 2.46^{+0.16}_{-0.06} $ & $ 2.336 ^{+0.002}_{-0.003} $ & $ 117.4^{+0.6}_{-0.4} $ & $ 1.240 (1040) $ & $ 0.42 \pm 0.04 $   \\

 IRAS 13349+2438${}^{e}$     & $ 7.89^{+0.10}_{-0.07} $ & $1.24^{+0.05}_{-0.00} $ & $ 38.0^{+0.5}_{-5.8} $ & $ 5.0^{+0.2}_{-0.1} $ & $2.190^{+0.003}_{-0.003} $ & $ 820^{+20}_{-10} $ & $ 1.198 (664) $ & $ 0.67 \pm 0.23 $  \\

 ARK 564${}^{e}$  & $7.9 ^{+0.2}_{-0.1} $ & $ 1.34^{+0.04}_{-0.04} $ & $ 58.2^{+0.7}_{-2.4} $ & $0.50^{+0.02}_{-0.01}$ & $2.479^{+0.005}_{-0.002} $ & $ 3120^{+100}_{-50} $ & $ 1.197 (666) $ & $ 0.72 \pm 0.07 $   \\

MRK 1044 & $ 9.3^{+0.3}_{-3.2} $ & $ 1.2^{+0.2}_{-0.0} $ & $ 57^{+3}_{-15} $ & $ 1.6^{+0.3}_{-0.3} $ & $ 2.28^{+0.02}_{-0.03} $ & $ 730^{+200}_{-120} $ & $ 1.081 (530) $ & $ 0.58 \pm 0.23 $  \\

E 1346+266${}^{d}$ & $ 1.3^{+8.7}_{-0.9} $ & $ 1^{+24}_{-0} $ & $ 90^{+0}_{-34} $ & $0.4 ^{+0.4}_{ -0.2} $ & $2.6^{+0.2}_{-0.1} $ & $ 2600^{+2970}_{-1080} $ & $ 1.068 (229) $ &  $ 0.46 \pm 0.16 $ \\

MRK 0359${}^{b}$ & $ 4.2^{+1.1}_{-0.7} $ & $ 1.2^{+0.6}_{-0.0} $ & $ 35^{+19}_{-10} $ & $ 0.6^{+0.2}_{-0.1} $ & $ 1.49^{+0.06}_{-0.06} $ & $ 690^{+110}_{-60} $ & $ 1.117 (527) $ & 1 \\

PHL 1092${}^{b,c}$& $10.0^{+0.0}_{-3.3} $&$ 1.24^{+0.07}_{-0.0} $&$ 69^{+1}_{-6} $&$ 1.0^{+0.1}_{-0.1} $&$ 2.54^{+0.01}_{-0.02} $&$ 1580^{+310}_{-60} $ & $ 1.237 (180) $ & 1 \\

RE J1034+396${}^{b}$ & $ 10.0^{+0.0}_{-1.6} $ & $ 1.2^{+0.1}_{-0.0} $ & $ 61^{+2}_{-5} $ & $ 0.37^{+0.03}_{-0.03} $ & $ 2.586^{+0.007}_{-0.007} $ & $ 3000^{+160}_{-890} $ & $ 1.081 (282) $ & 1 \\

 PKS 0558-504${}^{e}$  & $10.0^{+0.0}_{-1.4} $ & $ 1.24^{+0.02}_{-0.00} $ & $ 84.9^{+0.8}_{-0.3} $ & $ 0.84^{+0.05}_{-0.05} $ & $ 2.318^{+0.005}_{-0.007} $ & $ 1010^{+60}_{-40} $ & $ 1.063 (371) $ & $ 0.59 \pm 0.10 $  \\

 MRK 0766${}^{e}$  & $ 10.0^{+0.0}_{-1.2} $ & $ 1.7^{+0.3}_{-0.3} $ & $ 46^{+3}_{-6} $ & $ 1.3^{+0.4}_{-0.3} $ & $ 2.086^{+0.009}_{-0.006} $ & $ 370^{+40}_{-30} $ & $ 1.119 (952) $ & $ 0.34 \pm 0.09 $ & \\

 MRK 0586${}^{c}$  & $ 4.2   ^{+1.2}_{   -0.4} $ & $  1.4   ^{+0.4}_{-0.2} $ & $  55   ^{+9}_{   -10} $ & $   1.1  ^{ +0.2}_{  -0.2} $ & $  2.39   ^{+0.03}_{ -0.02} $ & $ 1300  ^{+130}_{  -100} $ & $   1.007 (523)  $ & $ 0.59 \pm 0.15 $   \\

TON S180 & $ 9.9^{+0.1}_{-0.6} $ & $ 1.235^{+0.001}_{-0.000} $ & $ 61^{+1}_{-3} $ & $ 0.69^{+0.06}_{-0.05} $ & $ 2.39^{+0.01}_{-0.01} $ & $ 3460^{+390}_{-390} $ & $ 1.396 (778) $ & $ 0.80 \pm 0.16 $  \\

\hline
\end{tabular}
\\
${}^{a}$ The fit includes redshifted cold absorption with an equivalent hydrogen column of  $ 10^{+2}_{-1} \times 10^{20}$~cm$^{-2}$. ${}^{b}$ The fit is entirely reflection dominated and includes no power law component. ${}^{c}$ Adding non-redshifted (local to our Galaxy) edges improves the fit, see Section \ref{notes_section}. ${}^{d}$ The model based on a non-rotating black hole has a similar goodness of fit, see Table \ref{nonrotating_fit_table}. ${}^{e}$ The fit includes warm absorption edges or a narrow Gaussian line at 6.4~keV, with parameters listed in Table \ref{disc_edges_table}.

\end{table*}

\begin{table*}
\caption{Redshifted absorption edges and narrow 6.4~keV Gaussian emission lines included in the relativistically-blurred disc fits in Table \ref{disc_fit_table}. Energy is the energy of the absorption edge, $\tau$ is the optical depth and Width is the equivalent width of the Gaussian line. Energies given without associated errors are fixed at 0.74 or 0.87~keV, the energy of the \textsc{O\thinspace VII} and \textsc{O\thinspace VIII} K absorption edges. Blank spaces indicate no feature is present.}\label{disc_edges_table}
\begin{tabular}{llllll}
\hline
Source & Energy (keV) & $\tau$ & Energy (keV) & $\tau$ & Width (eV)\\
\hline
PG 1307+085 & 0.74 & $ 0.6^{+0.2}_{-0.2} $ \\
PG 1309+355& 0.74 & $ 0.3^{+0.2}_{-0.1} $ & 0.87 & $ 0.2^{+0.1}_{-0.2} $ &\\
NGC 4051& 0.74 & $0.153^{+0.005}_{-0.007} $ & $ 0.463^{+0.007}_{-0.009} $ & $ 0.092^{+0.006}_{-0.017} $  & $90^{+20}_{-20} $\\
IRAS 13349+2438&  $0.72^{+0.20}_{-0.02} $ & $ 0.24^{+0.04}_{-0.02} $ & \\
ARK 564&$0.68^{+0.53}_{-0.01} $ & $0.10^{+0.01}_{-0.01} $ & \\
PKS 0558-504&  $ 0.72^{+0.02}_{-0.01} $ & $ 0.37^{+0.03}_{-0.03} $ &\\
MRK 0766& $ 0.708^{+0.005}_{-0.003} $ & $ 0.40^{+0.02}_{-0.01} $ & & & $50^{+40}_{-40} $  \\
\hline
\end{tabular}
\\
\end{table*}

\begin{table*}
\caption{Photoionized disc fits to the sources where the relativistic blurring is calculated assuming a non-rotating black hole. Column labels are as Table \ref{disc_fit_table}}\label{nonrotating_fit_table}
\begin{tabular}{llllllll}
\hline
Source & Em. Ind. &  $R_{in}$ & $i$ &  Fe & $\Gamma$ &  $\xi$ & $\chi^{2}_{\nu}$ (d.o.f.) \\
\hline
PG 1244+026 & $5.0^{+5.0}_{-2.1} $ & $ 6.0^{+4.5}_{-0.0} $ & $ 25.4^{+9.6}_{-22.2} $ &  $ 1.21^{+0.40}_{-0.23} $ & $ 2.44^{+0.03}_{-0.04} $ & $ 10000^{+0.0}_{-1970} $ & $ 1.108 (292) $ \\
E 1346+266 & $ 2.2^{+7.8}_{-2.2} $ & $ 6.0^{+6.6}_{-0.0} $ & $ 88.7^{+1.3}_{-42.1} $ & $ 0.45^{+0.28}_{-0.10} $ & $ 2.6^{+0.2}_{-0.1} $ & $ 2710^{+2090}_{-870} $ & $ 1.070 (229) $ \\
\hline
\end{tabular}
\\
\end{table*}

\section{Notes on individual sources}\label{notes_section}
\subsection{PG 0050+124}
PG 0050+124 (also known as I Zw 001) has a very weak soft excess. When the simple fit is used without a black body component the fit is only slightly worse: $\chi^{2}_{\nu}$ of 1.275 for 930 degrees of freedom, compared to a $\chi^{2}_{\nu}$ of 1.270 for 928 degrees of freedom. There is a $\Delta \chi^{2}$ of only 7 for 2 fewer degrees of freedom when the black body component is added to the fit, so it is barely significant.\\

\subsection{TON S180}
TON S180 has an unusual soft excess which Vaughan et al. (2002) successfully modelled as a power law rather than a black body (see Figure \ref{soft_excess_figure}). Neither of our models give an acceptable fit to the data, due to strong residuals in the \texttt{pn} spectrum below 0.5~keV (also present in the \texttt{MOS} spectrum, although of a slightly different shape). Vaughan et al. do not fit the \texttt{pn} spectrum below 0.5~keV, so their paper does not investigate the origin of these residuals. A disc reflection model with the Galactic cold absorption column fixed at zero (rather than $1.55 \times 10^{20}$~cm$^{-2}$, from Table \ref{sources_table}) can account for the shape of these residuals, with a $\chi^{2}_{\nu}$ of 1.066 for 778 degrees of freedom. A simple model fit with no Galactic absorption is still inadequate, with a $\chi^{2}_{\nu}$ of 1.264 for 782 degrees of freedom. These fits are unlikely to be physical, the Galactic column is rarely lower than $\sim 5 \times 10^{19}$~cm$^{-2}$. The residuals are instead likely to be due to some curvature of the soft excess not addressed by our models.\\

\subsection{PG 1404+226}
PG 1404+226 has been previously fit with the disc reflection model in Crummy et al. (2005). PG 1404+226 is strongly variable, and when the simple model is used it appears to have an absorption feature in the low state which disappears in the high state (see also Dasgupta et al. 2005). Crummy et al. show that the disc reflection model can reproduce both the high and low spectral states without the need for variable absorption.\\

\subsection{NGC 4051}
\begin{figure*}
\includegraphics[angle=270, width=0.45\textwidth]{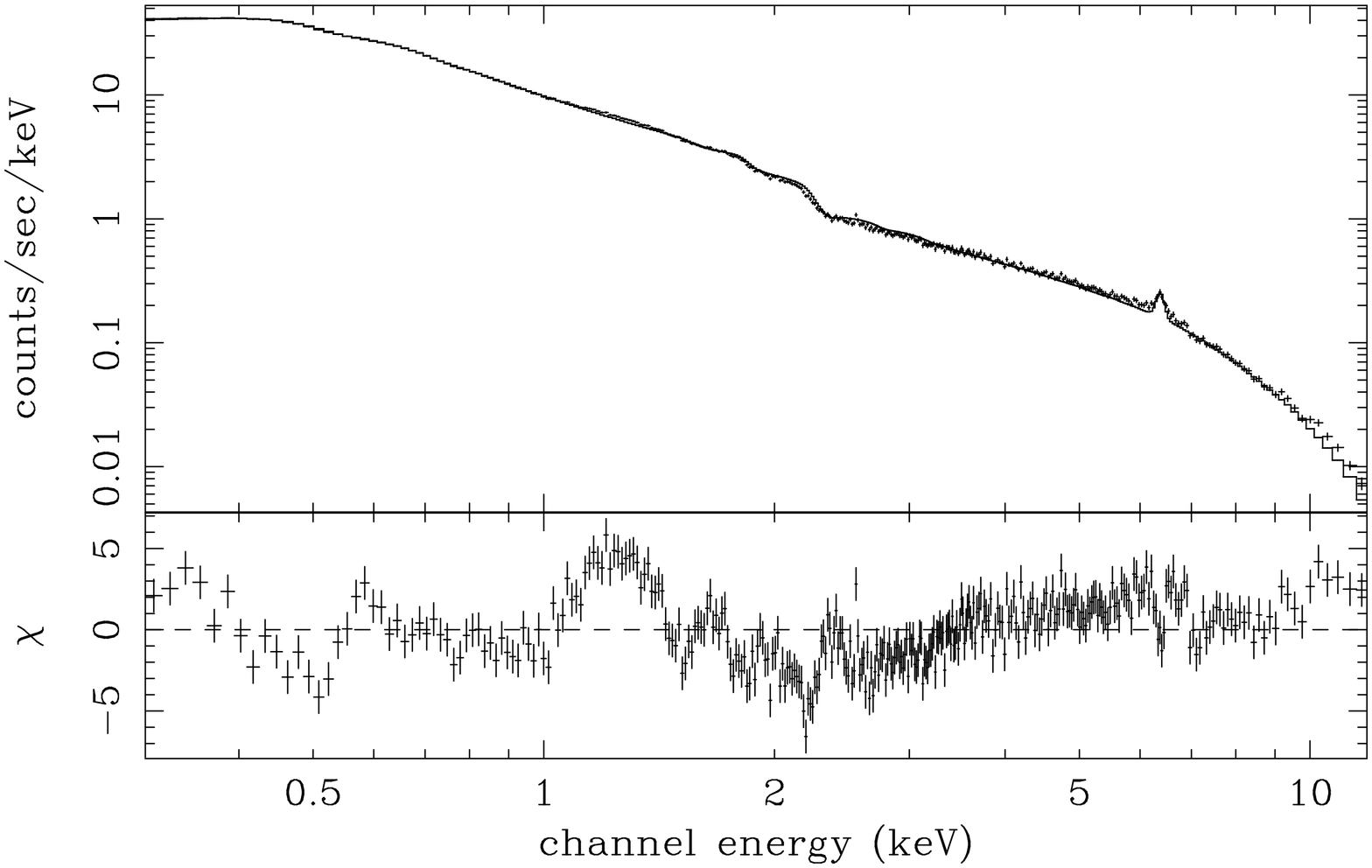}
\hspace{11mm}
\includegraphics[angle=270, width=0.45\textwidth]{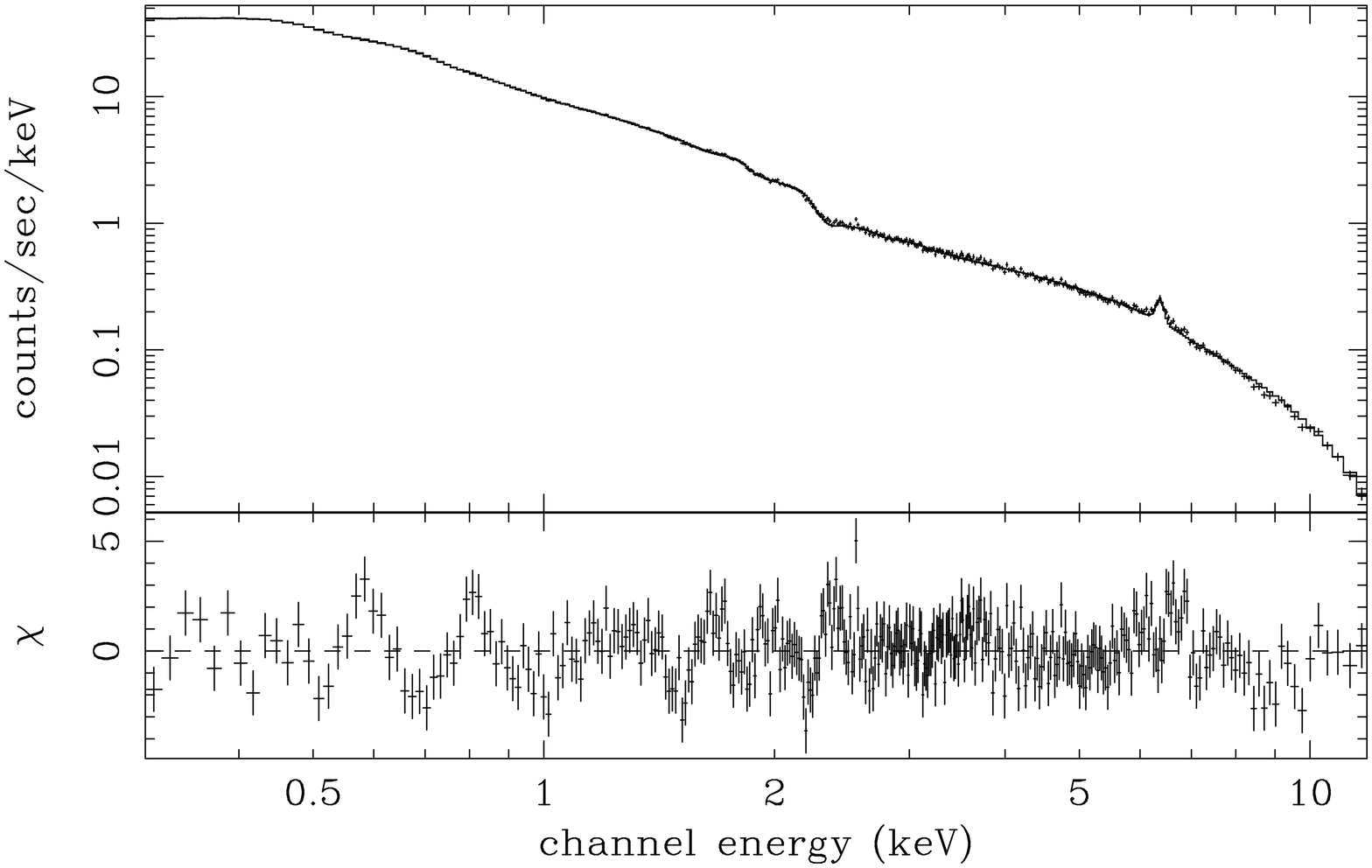}
\caption{The simple model (left) and relativistically-blurred photoionized disc reflection model (right) fits and residuals to NGC 4051, including Gaussian lines and absorption edges as stated in Tables \ref{phenom_fit_table}, \ref{disc_fit_table} and \ref{disc_edges_table}). The disc reflection model is clearly a superior fit, and follows the shape of the continuum very well, despite the high $\chi^{2}_{\nu}$. The spectra have been re-binned to show the shape of the residuals more clearly.}
\label{ngc4051_spectrum}
\end{figure*}

The observation of NGC 4051 we analyse is of very high quality, and shows some features not observable in any of the other sources. It is clear from Fig. \ref{ngc4051_spectrum} that the disc reflection model follows the shape of the continuum very well, but there are several small features which contribute to the fairly high $\chi^{2}_{\nu}$ of 1.240 for 1040 degrees of freedom. Some of these features may be calibration problems, such as around the \textit{XMM-Newton} instrumental Au M edge at 2.3~keV, and some may be due to small amounts of absorption or emission. However it is also possible that some of the approximations made in the disc reflection model are invalid when analysing observations of this detail, see Section \ref{discussion_section}. A more detailed analysis of NGC 4051 and its variability using the disc reflection model is in Ponti et al. (in prep.).\\

\begin{figure}
\begin{center}
\includegraphics[angle=270, width=0.45\textwidth]{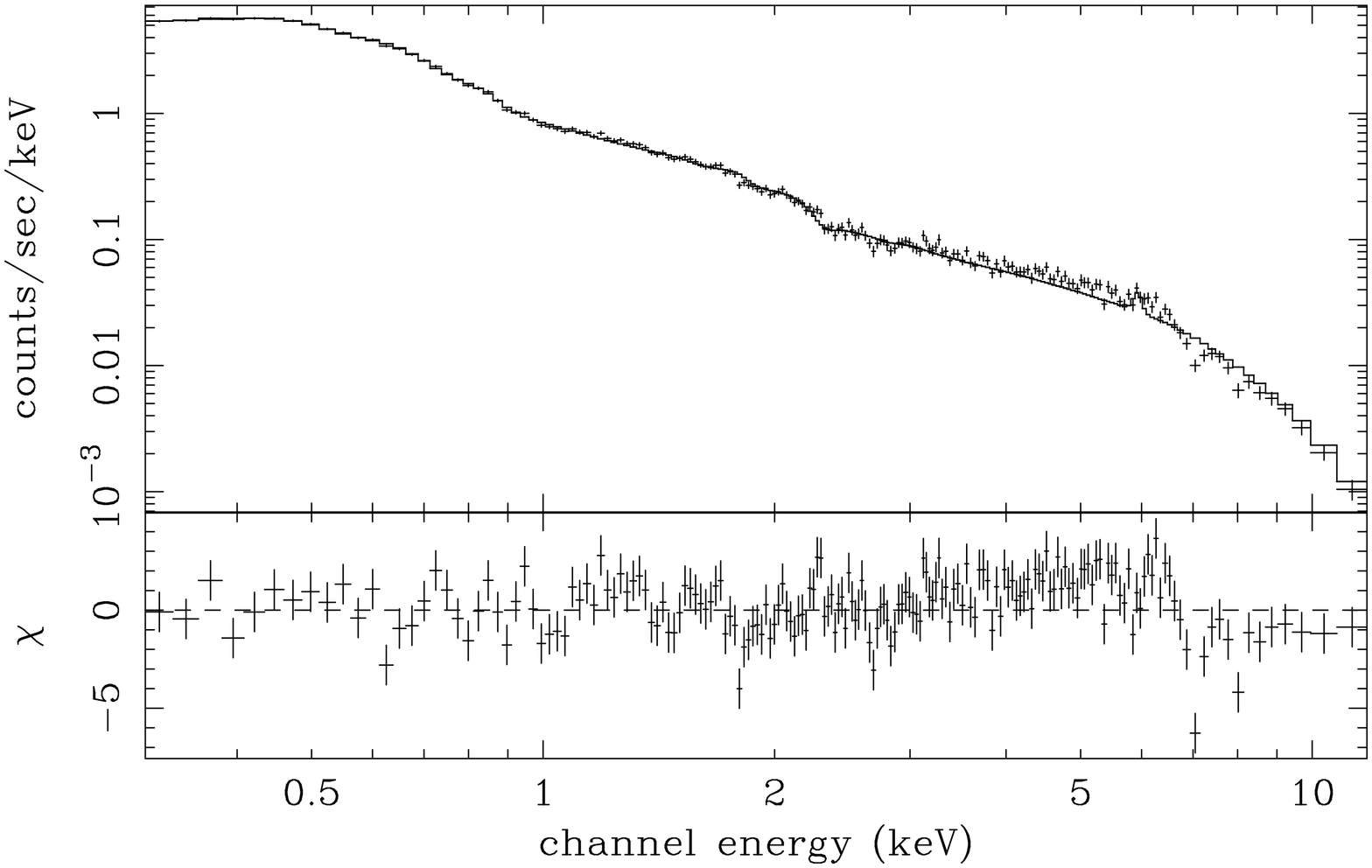}
\includegraphics[angle=270, width=0.45\textwidth]{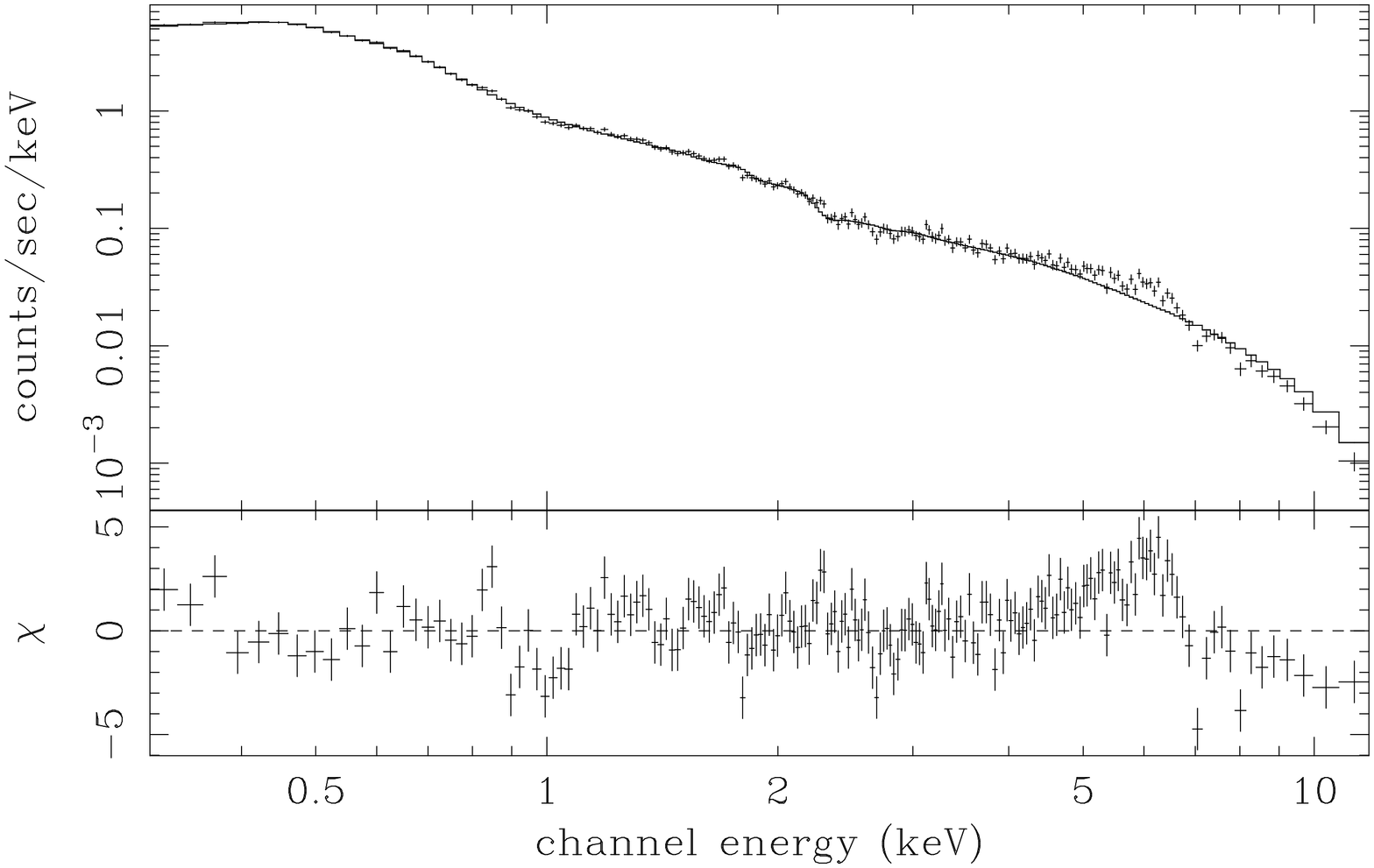}
\includegraphics[angle=270, width=0.45\textwidth]{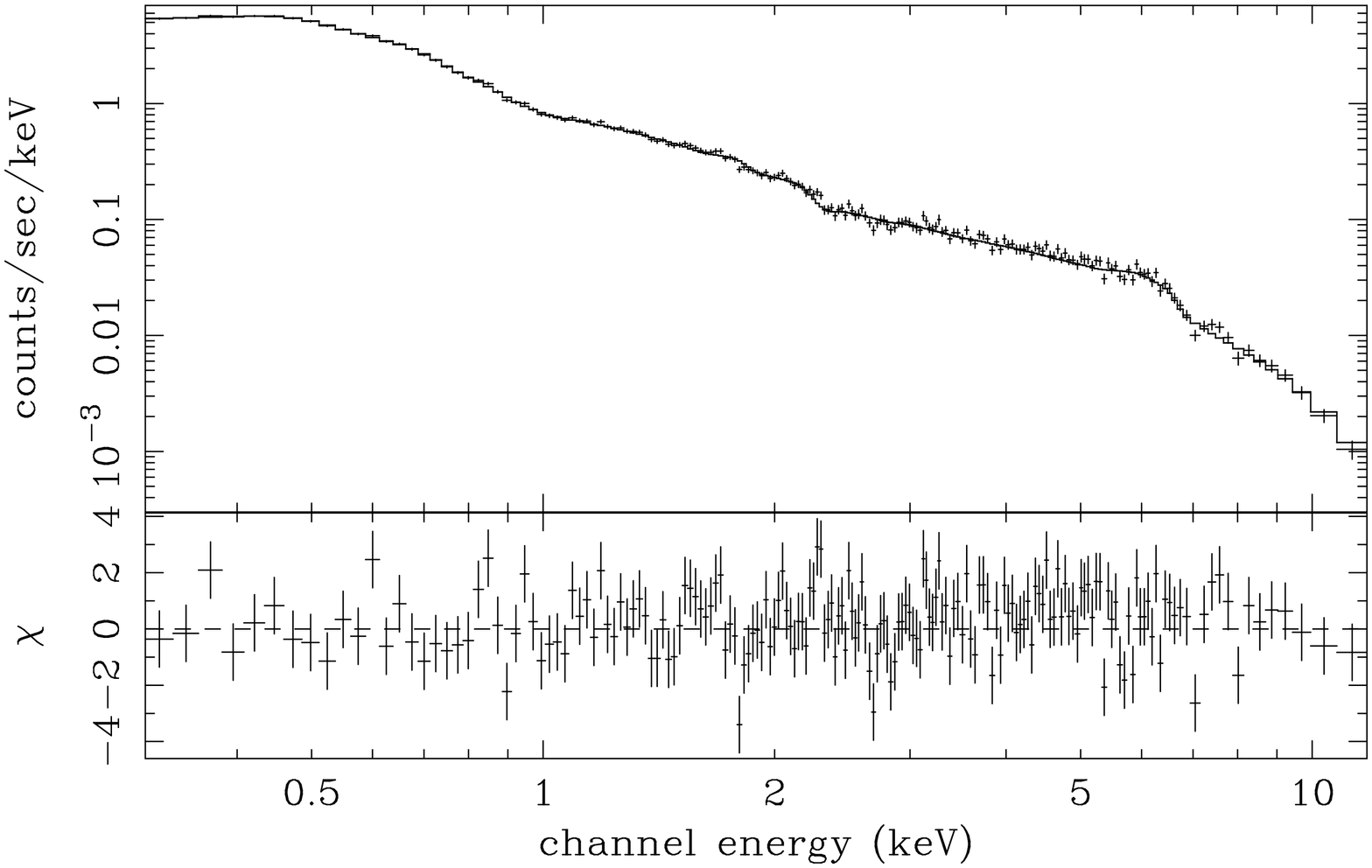}
\includegraphics[angle=270, width=0.45\textwidth]{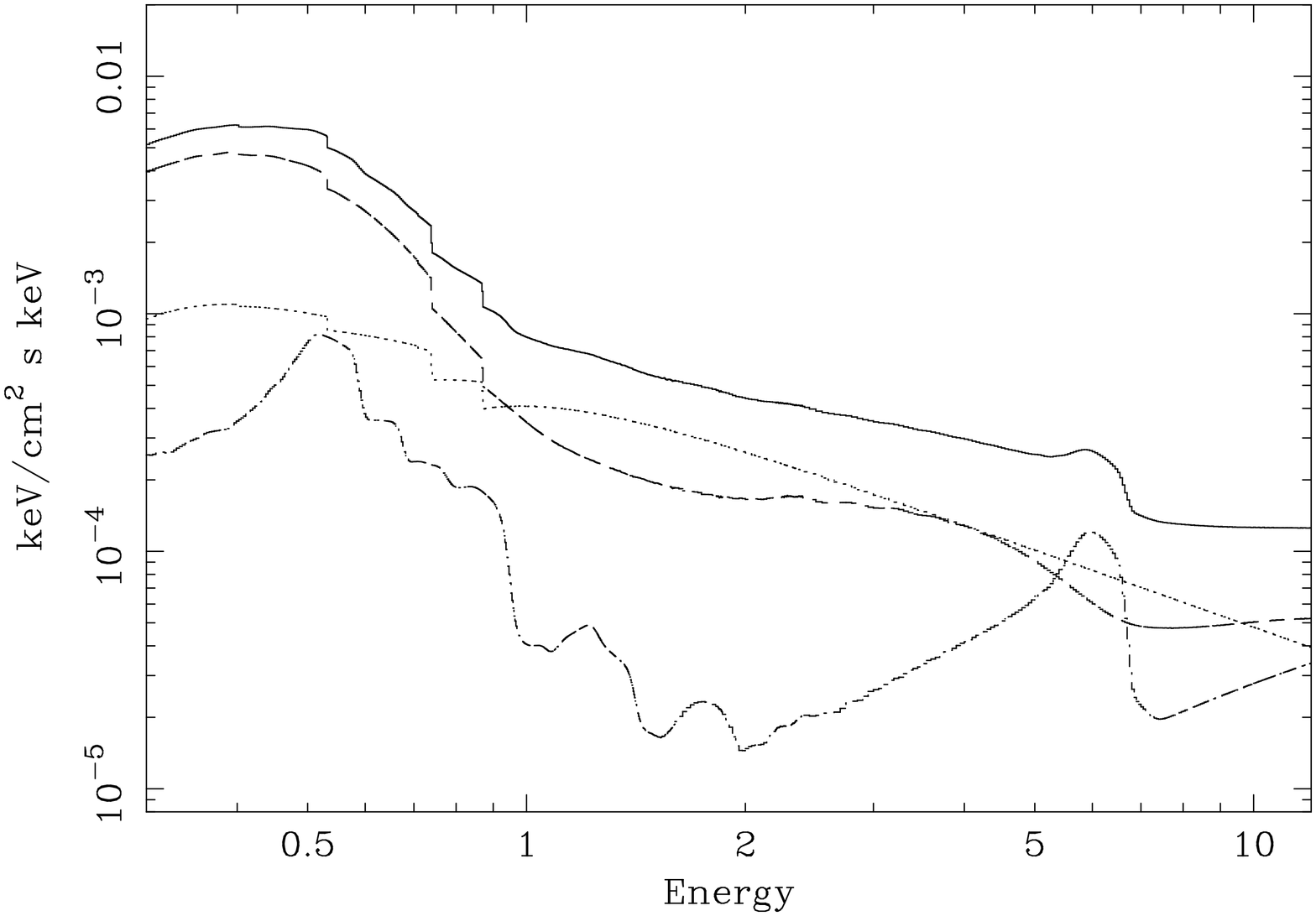}
\caption{From top to bottom the simple model, disc reflection model and double relection disc reflection model fits and residuals to PG 1211+143, and a plot of the double reflection disc reflection model (the solid line is the total model, the dotted line is the power law and the dashed and dash-dotted lines are the reflection components). PG 1211+143 has some structure in the 4 -- 8~keV region, visible in the top two spectra, that is best explained by adding a second reflection component with an inner radius of 3 gravitational radii, as shown in the bottom panels. See Section \ref{notes_section} for a discussion of this feature. The spectra have been re-binned for display.}
\label{1211}
\end{center}
\end{figure}

\subsection{E 1346+266}
E 1346+266 is unusual in our sample as it has a high redshift, z = 0.915. This enables us to investigate the effectiveness of the disc reflection model at source energies up to 23~keV. The simple model, the disc reflection model and the version of the disc reflection model with a non-rotating black hole are all roughly equal in quality of fit to the source, so whilst the disc reflection model has proved itself to work at least as well as the simple model in this energy range it is difficult to say anything definite about the source. The disc reflection model essentially works by fitting the soft excess and any excess iron emission. The soft excess is less visible in high redshift AGN due to being both shifted out of the instrument band and strongly absorbed. The disc reflection model will therefore be most useful in high redshift sources which have excesses in the continuum around 6.4~keV due to iron, although it is able to fit sources where this is not the case.\\

\subsection{PG 1211+143}
PG 1211+143 has a feature in the 4 -- 8~keV region which is not predicted by either model, see Fig. \ref{1211}. The shape of the residuals to the fits suggest two possibilities, an absorption edge at $\sim$7~keV or a broad Laor iron line. We investigated the latter possibility by adding a Laor line at 6.4~keV (in the source frame) to our reflection model, with the inclination, emissivity index and outer radius fixed to the same values as our Laor blurring, allowing only the inner radius and normalisation to vary. We found this improved the fit very significantly, to a $\chi^{2}_{\nu}$ of 1.053 for 932 degrees of freedom. However, this is unphysical as any iron line would also have other reflected lines associated with it. We therefore fit PG 1211+143 with two reflection components, both illuminated by the same power law and at the same inclination, with the same iron abundance, but allowing all other parameters to vary. We found that this improves the fit further, to a $\chi^{2}_{\nu}$ of 1.030 for 930 degrees of freedom, with illuminating power law index $2.07^{+0.02}_{-0.01}$, non-redshifted edges (i.e. potentially local to our Galaxy) at 0.74 and 0.87~keV with depths of $0.28^{+0.03}_{-0.09}$ and $0.24^{+0.04}_{-0.04}$ respectively, disc inclination $47.6^{+1.1}_{-0.4}$ degrees, iron abundance $6.4^{+0.4}_{-1.2}$ times solar, emissivity indices $9.2^{+0.1}_{-0.3}$ and $4.8^{+1.3}_{-0.6}$, inner radii $1.31^{+0.06}_{-0.07}$ and $3.0^{+0.2}_{-0.7}$ gravitational radii, and ionization parameters $610^{+140}_{-40}$ and $29^{+1}_{-10}$~erg cm s$^{-1}$. Where we give two parameters the second refers to the outer disc component, with an inner radius of 3 gravitational radii.\\
The fit parameters are not consistent with the fit in Table \ref{disc_fit_table}, but given that the fit has improved so much ($\Delta \chi^{2}_{\nu}$ of 235 for 7 degrees of freedom) that is entirely expected; in particular the iron abundance has reduced from 10 to 5 since a single ionized disc is no longer trying to fit both the large iron line and the soft excess. This model is somewhat similar to the work of Sobolewska \& Done (2004), who achieve a similar goodness of fit to PG 1211+143 with three unblurred reflection components. Taking account of relativity by adding relativistic blurring removes the need for one of their reflectors, we instead have a power law. Our model is also close to that of Fabian et al. (2002) who use three blurred reflectors to model 1H 0707-495.\\
We also investigated the alternative possibility already reported by Pounds et al. (2003) which is an absorption edge at $\sim$7~keV. We added a redshifted edge at 7~keV with energy and depth as free parameters to our reflection model, and found that it too improved the fit significantly. This model has a $\chi^{2}_{\nu}$ of 1.122 for 934 degrees of freedom, with emissivity index $6.48^{+0.05}_{-0.05}$, inner radius $1.24^{+0.05}_{-0.00}$ gravitational radii, inclination $35.3^{+0.4}_{-0.4}$ degrees, iron abundance $4.73^{+0.09}_{-0.10}$ times solar, illuminating spectral index $1.88^{+0.01}_{-0.01}$, ionization parameter $300^{+3}_{-3}$~erg cm s$^{-1}$, edge energy $7.44^{+0.03}_{-0.10}$~keV and edge depth $0.84^{+0.10}_{-0.08}$. A plausible way to create the absorption edge is a highly ionized outflow (Pounds et al. 2003). We find that adding an edge at $\sim$7.3~keV to the disc reflection model is an improvement, but that adding a second reflection component at 6.4~keV gives an even better fit. Physically, since this second component has a larger inner radius, it may represent a a hot spot or ring of emission excited by e.g. flares from the magnetic fields.\\

\subsection{PHL 1092}
PHL 1092, when fit with the disc reflection model, appears to have a non-redshifted edge at 0.87~keV. The fit parameters when this edge is included are a $\chi^{2}_{\nu}$ of 1.208 for 179 degrees of freedom (compared to 1.237 for 180 degrees of freedom without the edge), edge depth $0.3^{+0.1}_{-0.2}$, emissivity index $9.9^{+0.1}_{-2.2}$, inner radius $1.24^{+0.08}_{-0.0}$ gravitational radii, inclination $69.4^{+2.5}_{-6.2}$ degrees, iron abundance $1.00^{+0.06}_{-0.05}$ times solar, illuminating spectral index $2.53^{+0.01}_{-0.01}$ and ionization parameter $1460^{+350}_{-50}$~erg cm s$^{-1}$.\\
PHL 1092 has been fit by Gallo et al. (2004a), who use the radiative efficiency formula of Fabian (1979) to conclude that the variability of the source is inconsistent with a non-rotating black hole, whilst being consistent with a maximally rotating black hole. They suggest that the source is reflection dominated, and that light bending may be in operation. This is consistent with our best fit for PHL 1092, which is pure reflection without a power law component.

\subsection{MRK 0586}
MRK 0586, when fit with the disc reflection model, appears to have a non-redshifted edge at 0.87~keV. The fit parameters when this edge is included are a $\chi^{2}_{\nu}$ of 0.977 for 522 degrees of freedom (compared to 1.007 for 523 degrees of freedom without the edge), edge depth $0.15^{+0.05}_{-0.05}$, emissivity index $4.8^{+0.5}_{-1.1}$, inner radius $1.24^{+0.60}_{-0.00}$ gravitational radii, inclination $51.3^{+9.7}_{-8.0}$ degrees, iron abundance $0.75^{+0.09}_{-0.10}$ times solar, illuminating spectral index $2.33^{+0.02}_{-0.02}$ and ionization parameter $1710^{+420}_{-200}$~erg cm s$^{-1}$.\\
MRK 0586 has been previously investigated by Gallo et al. (2004b) who find a feature consistent with a highly broadened iron line emitted from the inner regions of an accretion disc around a Kerr black hole. Our disc fit, where the fit is dominated\footnote{The majority of the photon flux is in the soft energy band, and fitting just the soft band recovers parameters consistent with fitting over the entire available energy range.} by the soft excess rather than the weak iron line, supports their hypothesis; we find an inclination ($55^{+9}_{-10}$ degrees) consistent with theirs ($37^{+9}_{-14}$) although we have fit the emissivity index and find $4.2^{+1.2}_{-0.4}$ where they fix it at 3.\\

\section{Discussion}\label{discussion_section}
The relativistically-blurred photoionized disc reflection model is clearly a better fit in terms of reduced chi-squared. 25 of the 34 sources show a significant ($\Delta \chi^{2}_{\nu} >$ 2.7 per degree of freedom; note that this does not correspond to a 90 per cent probability as the $\chi^{2}$ distribution is not calibrated between models) improvement when moving from the simple model to the disc reflection model, with 6 sources showing a significant worsening and the remaining 3 sources inconclusive. The disc reflection model is a good fit to the sources ($\chi^{2}_{\nu} <$ 1.1) in 22 of the sources, an adequate fit ($\chi^{2}_{\nu} <$ 1.3) in 10 cases, and a poor fit ($\chi^{2}_{\nu} >$ 1.3) to TON S180, compared to 19 good fits, 13 adequate fits and 2 poor fits with the simple model. The sources which are fit worse by the disc reflection model are PG 1211+143, PG 1309+355, IRAS 13349+2438, PHL 1092, RE J1034+396 and TON S180. PG 1211+143 can be well fit by the disc reflection model if a second reflection component is added (Section \ref{notes_section}), the other sources show no obvious commonalities, being both high and low quality observations with fit parameters spread out over a similar range to the entire sample. Several of these sources show small excesses around 6~keV with a drop at 7 -- 8~keV, which may be due to extra iron emission like PG 1211+143, however PG 1309+355 does not, and this feature is not uncommon throughout the entire sample.\\
The disc reflection model accounts naturally for the shape of the continuum in all the sources. The model also naturally accounts for features which would be interpreted as absorption edges in the simple model -- 17 sources appear to have absorption edges when the simple model is used, reducing to 7 sources using the disc reflection model. We therefore conclude that the relativistically-blurred photoionized disc reflection model reproduces the spectra of our sources better than the simple model, and should be a primary tool in the investigation of X-ray observations of AGN.\\
\subsection{Warm absorption}
Only a few sources may show warm absorption edges local to our Galaxy. When using the simple model we find only one source where a non-redshifted (i.e. local) edge improves the fit, and we eliminate that as a possibility based on a variability analysis in a previous paper (see PG 1404+226 in Section \ref{notes_section}). When using the disc reflection model we find three sources that may exhibit non-redshifted edges (PG 1211+143, PHL 1092 and MRK 0586), none of which have warm absorption edges local to the sources. Whilst our analysis is not optimised to detecting these features, we can conclude that strong non-redshifted edges are rare. A further point to note is that the disc reflection model finds more of them than the simple model, which is the opposite of the situation with edges local to the sources. Warm absorption edges local to the source appear to be very common when the simple model is used, but many of the features that seem to be edges can be explained by the shape of the disc reflection model. Many features previously identified as absorption edges may instead be due to blurred reflection from an ionized disc; high resolution observations and variability studies offer methods to distinguish between the two cases.\\
\subsection{Inclinations}
\begin{figure}
\begin{center}\includegraphics[angle=0, width=0.45\textwidth]{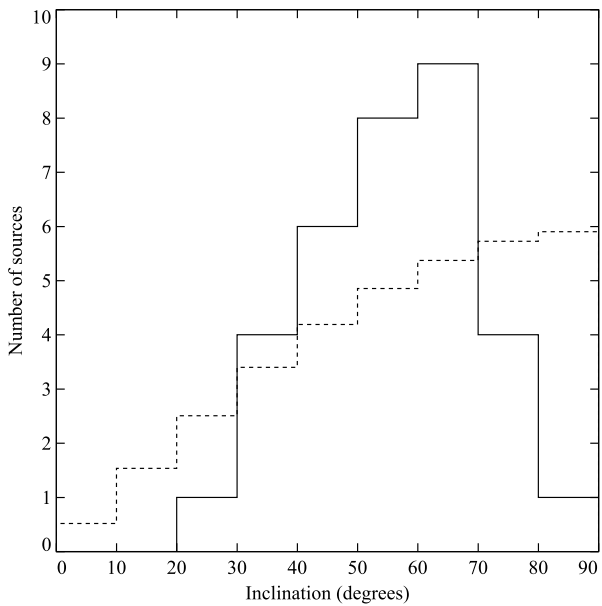}
\caption{A histogram showing the measured inclinations of the sources (solid line) with the random distribution over-plotted (dotted line). There is a deficit of sources at high inclinations compared to the random distribution, which supports the AGN unified model.}
\label{inclination_figure}
\end{center}
\end{figure}

\begin{figure}
\begin{center}\includegraphics[angle=0, width=0.45\textwidth]{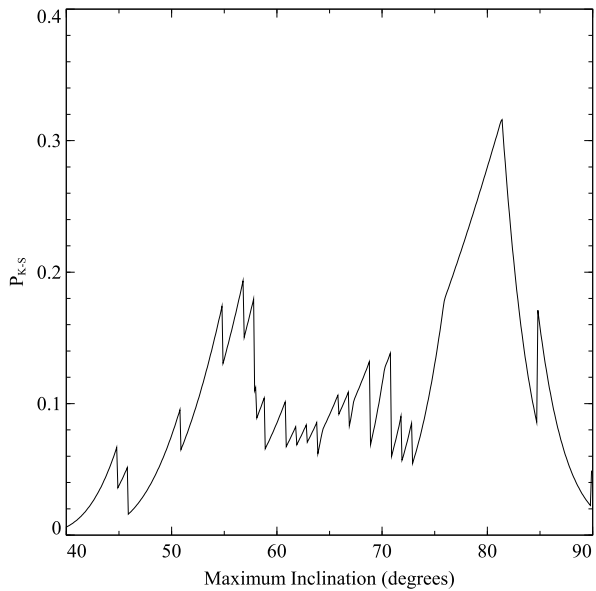}
\caption{The Kolmogorov-Smirnov probability ($P_{K-S}$) that the measured inclinations are consistent with a random distribution over the range $0^{\circ}$ to a maximum inclination (x-axis). The data are inconsistent with being random over the entire $90^{\circ}$ range, though they are reasonably consistent with being random over $0^{\circ}$ -- $80^{\circ}$, giving support to an AGN unified model with a with a torus that obscures discs of inclinations above $80^{\circ}$.}
\label{kolmogorov_figure}
\end{center}
\end{figure}

The model parameters we measure give us information about our sample. The ranges of inclinations are plotted in Fig. \ref{inclination_figure}. It should be noted that these inclinations are rather model dependent, and that the formal errors from \texttt{xspec} fitting are likely to be underestimates. The inclination of the accretion disc to the line of sight spans the a range from 22$^{\circ}$ -- 90$^{\circ}$, though three sources have a inclinations consistent with 0$^{\circ}$ within the errors (see Table \ref{disc_fit_table}). A random distribution in inclination would be proportional to $\sin(i)$, which is not consistent with our measurements. A Kolmogorov-Smirnov (K-S) test measures the probability that a set of data is consistent with a given distribution. A K-S test against a random distribution to our data, neglecting the measured uncertainties in inclination, gives a probability of 0.06. Under the unified model of active galactic nuclei (Antonucci 1985) they would be confined to a range in inclination, $P(i) \propto \sin(i);$ $i < \alpha$ where $90 - \alpha$ is the angular extent of the torus above the plane of the disc in degrees. The unified model can explain the apparent deficit in discs with an inclination above $70^{\circ}$, though it does not explain the sources which we do measure with high inclinations, nor the apparent deficit below $30^{\circ}$. We performed a series of K-S tests (in \textsc{octave 2.1.60}) using our simple unified model distribution with cut-offs $\alpha$ ranging from $0^{\circ}$ to $50^{\circ}$, i.e. over a range in maximum inclinations of $40^{\circ}$ -- $90^{\circ}$, discarding any measured inclinations above the cut-off. The results are shown in Fig. \ref{kolmogorov_figure}. We found a maximum K-S test probability of $\sim$ 0.34 using an $81^{\circ}$ cut-off, and probabilities of $\sim$ 0.1 with cut-offs over the range $50^{\circ}$ -- $85^{\circ}$. The data are incompatible with being randomly distributed over the entire $0^{\circ}$ -- $90^{\circ}$ range in inclination, but are somewhat consistent with being random over $0^{\circ}$ -- $80^{\circ}$. The deficit below $30^{\circ}$ may be partly explained by a selection effect; under the relativistic disc model sources emit most of their light along the disc and very little directly upwards, due to Doppler beaming. It is also possible that it is an artefact of our small sample size.\\
The emission from the relativistically-blurred photoionized disc reflection model is dominated by the innermost part of the accretion disc, if the inner accretion disc is not always coplanar with the torus (Bardeen \& Petterson 1975) then our high inclination sources can be explained; however, estimates of the time taken for a Bardeen-Petterson inner disc to align with the outer disc are $\lesssim 10^{6}$~years (Natarajan \& Armitage 1999) so Bardeen-Petterson discs should be rare. It is also possible that some of these type 1 AGN have a small or no torus.\\
\subsection{Black hole rotation}
For every source in our sample we found that using relativistic blurring based on a non-rotating black hole was a worse fit than a maximally rotating hole (Section \ref{results}), with only two sources where the fit is not significantly worse. This implies that rapidly rotating black holes dominate our sample. More information can be extracted by considering the measured inner radii of our sources. An accretion disc around a black hole is not expected to extend further in than the last stable orbit. The last stable orbit around a non-rotating (Schwarzschild) black hole is at 6 gravitational radii. All 34 sources have inner accretion disc radii consistent with being under 6 gravitational radii, and only two sources (PG 1244+026 and E 1346+266, the same sources that are equally consistent with a non-rotating fit) have an inner radius which may be over that limit. The last stable orbit around a maximally rotating (Kerr) black hole is at 1.235 gravitational radii (this requires a disc with angular momentum in the same direction as the black hole angular momentum). Almost all of the sources (29 of the 34) have inner radii consistent with being below 1.3 gravitational radii. This shows that the majority of black holes in (type 1) AGN are rapidly rotating, in agreement with theoretical predictions based on the fact that black holes gain most of their mass from accretion discs which tend to increase the angular momentum. Volonteri et al. (2005) predict a distribution of spins where 70 per cent of AGN black holes are maximally rotating, which agrees well with our measurements.\\
Black hole rotation has also been measured for more distant quasars by Streblyanska et al. (2005), who find an iron line described by a Laor profile with an inner radius of $\sim$3 gravitational radii in a combined spectrum of distant quasars from the Lockman Hole. This implies that the average distant AGN rotates rapidly.\\
It should be noted that the black hole rotation is not a free parameter in our analysis; we perform the fits assuming a maximally rotating black hole therefore we have not measured the rotation, merely found that many of the fits are statistically consistent (low $\chi^{2}_{\nu}$) with maximal rotation. An alternative possibility is that radiation from inside the last stable orbit (the plunging region) may be important. See Krolik \& Hawley (2002) for a discussion of the inner edges of accretion discs and Section 3.4 in Fabian \& Miniutti (2005) for comments on why the inner radius is likely to be a good estimator of black hole spin.\\
\subsection{Emissivity indices}
\begin{figure}
\begin{center}\includegraphics[angle=0, width=0.45\textwidth]{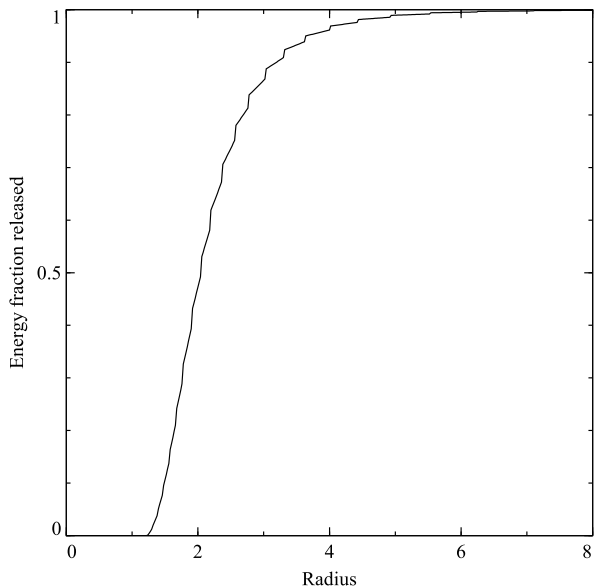}
\caption{A cumulative plot of the flux in a Laor line with parameters given in Section \ref{discussion_section} emitted as a function of radius. This was generated by varying the inner radius parameter and normalising to the flux emitted when the inner radius is 1.235 gravitational radii. The small step-like features in the plot are due to numerical inaccuracies in the model.}
\label{power_figure}
\end{center}
\end{figure}

The emissivity index of our sources tends to be very high, with 22 sources having an index over 6, 11 sources with an index between 3 and 6, and only one, E 1346+266, with an index below 3. The measurements for E 1346+266 are not strongly constrained as it is at a fairly high redshift (see Section \ref{notes_section}), and the data are consistent with a much higher index. For a typical source with an emissivity index of 7, inclination of 50$^{\circ}$, and inner radius of 1.235 gravitational radii an observer will detect the radial flux profile shown in Fig. \ref{power_figure}. Energy release is very concentrated towards the centre; 90 per cent of the flux is emitted within 2.0 gravitational radii. The energy observed is slightly less concentrated, with 90 per cent of the flux originating within 3.2 gravitational radii. This is different to the result of na\"{i}vely integrating the emissivity profile due to relativistic effects, however it is still clear that we are probing very close to the central black hole. This suggests that a study to compare measured variability timescales with emission region sizes from the disc reflection model could prove fruitful.\\
\subsection{Abundances}
The iron abundance measured from the sources is consistent with being between solar and $1/3$ solar for 28 of the 34 sources. Six sources have iron abundances which are definitely above solar. These fairly low abundances partially explain the lack of obvious broad iron lines. Care should be taken in drawing strong conclusions about metallicities from this as the model only allows for a varying iron abundance, the abundances of all other elements are fixed at solar. Furthermore, fits with the iron abundance fixed at a high value are still acceptable in some cases. However, these results encouragingly show that this technique is capable of measuring metallicities of accretion discs, and with higher quality data and a model which allows the measurement of abundances for more elements the question of the composition of AGN accretion discs could be answered by observing their X-ray emission.\\
\subsection{Spectral indices}
\begin{figure}
\begin{center}\includegraphics[angle=0, width=0.45\textwidth]{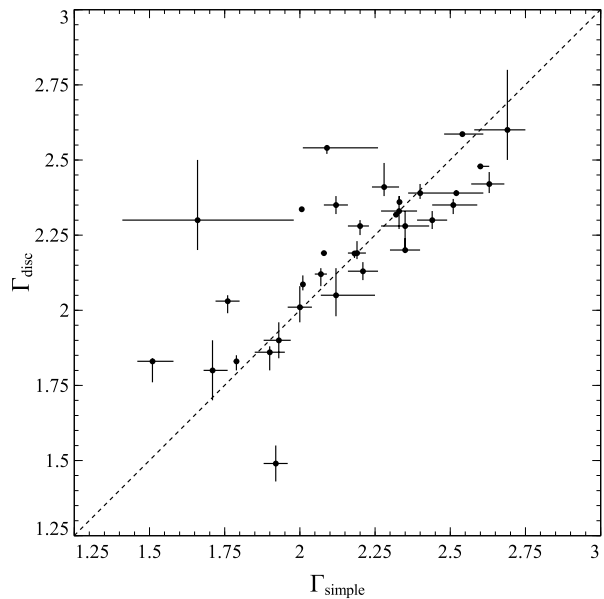}
\caption{A plot comparing spectral indices measured using both models. $\Gamma_{simple}$ is the spectral index from the simple model, $\Gamma_{disc}$ is the spectral index from the relativistically-blurred photoionized disc reflection model. The dotted line shows a 1:1 ratio. Most sources have a similar spectral index under both models, but the simple model has several outliers with a lower spectral index than expected.}
\label{indices_figure}
\end{center}
\end{figure}

\begin{figure}
\begin{center}
\includegraphics[angle=270, width=0.45\textwidth]{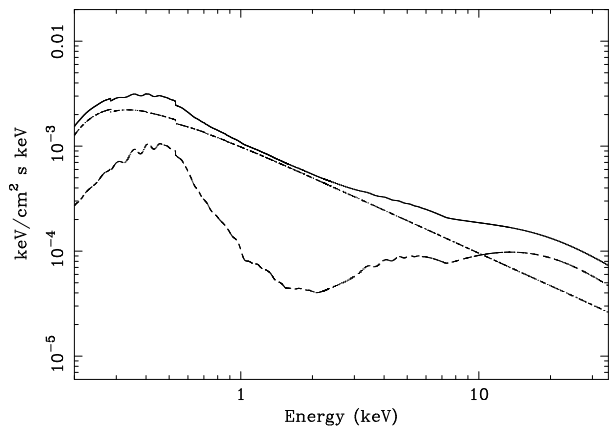}
\caption{The relativistically-blurred photoionized disc reflection model used to fit PG 1202+281, shown over an extended energy range. The dashed line is the reflection component, the dash-dotted line is the illuminating power law, and the solid line is the total model. The curvature at low energies is due to Galactic cold absorption. See Table \ref{disc_fit_table} for fit parameters. Note that the slope of the overall spectrum is significantly different from the slope of the power law component due to the high energy upturn in the reflection component (see Spectral indices in Section \ref{discussion_section}).}
\label{1202_emo_figure}
\end{center}
\end{figure}

The spectral index we measure from the sources using both models is shown in Fig. \ref{indices_figure}. Most of the sources have similar indices under both models, but there are several sources which have much lower spectral indices under the simple model (PG 1202+281, PG 1307+085, PG 1404+226, NGC 4051, and PHL 1092). PHL 1092 has no power law component in its spectra when fit with the disc reflection model, and a very strong soft excess, so the disc reflection model has a steeper illuminating continuum than the simple model. The other four sources have reflection components which increase in strength above $\sim$8~keV so the simple model, which effectively fits the entire high energy region with just a power law, will be flatter. See Fig. \ref{1202_emo_figure}. These results show that caution should be exercised when measuring the spectral index of a source if the intention is to relate it to a physical model; if disc reflection is present the spectral index will be due to the sum of more than one processes.\\

\subsection{The relativistically-blurred photoionized disc reflection model}\label{model_section}
The relativistically-blurred photoionized disc reflection model is an improvement in both goodness of fit and physical motivation over the simple model. However, in its present form there are still residuals to the fit in several of the sources. These residuals tend to occur in the 6 -- 8~keV region, and show a variety of features: some may be absorption edges or notches, some emission lines (narrow and broad). The disc reflection model is very smooth due to the relativistic blurring, it cannot produce a notch, though the model can explain features that look like absorption edges. The sharper features are therefore probably due to iron (and possibly other elements like nickel) emission and absorption, with the broader features due to either absorption from gas in the line of sight or iron emission from further out in the disc -- perhaps from regions excited by magnetic reconnection. It is possible that our current model of a geometrically flat accretion disc with a power law emissivity index does not accurately represent all the AGN studied.\\
The photoionized reflection spectra are generated by a model of illumination onto a slab of gas of constant density. This is unlikely to be the situation in a real AGN accretion disc, but is a necessary approximation to allow the calculation of a large grid of models such that fits can be performed with \texttt{xspec}. Previous studies have compared constant density reflection spectra to those which use a more physical density distribution (e.g. hydrostatic equilibrium, which still requires assumptions about the structure of the disc and other factors like the external coronal pressure). Ballantyne, Ross \& Fabian (2001) found that their hydrostatic equilibrium model could be approximated well by a ``diluted'' constant density model, where diluted means that the reflection fraction of the constant density model is smaller than the reflection fraction of the hydrostatic model it is fitting. This happens because an outer layer of the disc becomes highly ionized (a low density means a high ionization parameter) to the point where it produces no line features, and the lines form in a partially hidden layer below this. This may mean that the flux fractions we measure in Section \ref{fluxfrac_section} are too low, some of the emission we interpret as due to the illuminating power law might be reflection from a highly ionized surface layer.\\
An accretion disc is an uncontroversial component in a model of an AGN, but our model involves an accretion disc illuminated by a power law. This raises the question of the origin of the illuminating radiation. A corona of hot electrons above the accretion disc is often hypothesised, and can produce a power law by Comptonisation of UV disc photons (or bremsstrahlung or synchrotron emission). Alternatively a small hot-spot, either the base of a jet, shocks in a failed jet (Ghisellini, Haardt \& Matt 2004) or a region excited by a magnetic flare above the disc, can produce the necessary radiation. This model seems more likely in the context of our results -- we have several sources where the illuminating continuum is not visible to us (Table \ref{disc_fit_table}, Figure \ref{flux_frac_figure}) and this can be explained by the light bending effects of the central black hole -- photons emitted from sources above the disc tend to be bent towards it and very few escape to be observed, increasing the measured flux fraction by supressing the power law. Reflected radiation escapes the black hole more easily, it is Dopper beamed along the plane of the disc, away from the black hole. This is discussed in Miniutti et al. (2003) and Miniutti \& Fabian (2004). However, it should be noted that our results show no correlation between flux fraction and inclination, so we consider that Doppler beaming along the disc is not the dominant factor. This would also be the case if the illuminating hot spot rotates with the disc, sharing the same velocity and beaming; unlikely for jet based illumination but expected for magnetic flares above a disc or a corona. The consistently small inner radius is difficult to explain with magnetic flares. Alternatively, with a small power law emitting region it is possible that the power law radiation is obscured, although since 1 in 3 of the sources have no power law component and all sources have disc components the geometry of the absorber is difficult to imagine. Our favoured scenario is a small power law illuminating region above the central black hole, where light bending effects mean that the illuminating power law radiation can be deflected onto the disc and not be detected..\\
One further problem that the relativistically-blurred photoionized disc reflection model can solve is the general absence of strong iron lines in AGN spectra. An accretion disc illuminated by X-rays should produce an iron line. The disc reflection model shows that the iron line is present, but relativistically blurred to such an extent that it is hard to detect, see Figure \ref{nufnu_figure}.\\

\section{Conclusions}
We investigated a large sample of type 1 AGN X-ray spectra, and fit them with the relativistically-blurred photoionized disc reflection model of Ross \& Fabian (2005). This work has shown that:
\begin{itemize}
\item The relativistically-blurred photoionized disc reflection model fits the sample better than the conventional simple model. Since the disc reflection model is physically motivated, unlike the simple model, it can also give us direct information about the inner accretion disc.
\item The disc reflection model reproduces the continuum shape of all the AGN in the sample, in particular the soft excess is naturally explained by the model. The soft excess is composed of many broad lines mixed together.
\item The disc reflection model reproduces many features in the spectrum that could otherwise be interpreted as warm absorption edges.
\item The majority of black holes in the sample are strongly rotating, alternatively it is possible that radiation from the plunging region is important.
\item The inclinations measured from the sample span the range of 20$^{\circ}$ -- 90$^{\circ}$, so neither edge of the distribution is consistent with the simplest unified model. However, there is a deficit at inclinations $> 70^{\circ}$ so there is some evidence for torus obscuration.
\item The disc reflection model provides evidence that the iron abundances of black hole accretion discs tend to be solar or mildly sub-solar.
\end{itemize}
We have shown that the relativistically-blurred photoionized disc reflection model is an important tool in the study of AGN, and that taking account of the intrinsically relativistic nature of these sources not only provides information about the state of the system, but explains the apparent absence of iron line reflection from the disc and the shape and behaviour of the soft excess.

\section*{Acknowledgments}
J.C. is a UK PPARC funded PhD student. A.C.F. thanks the Royal Society for support. R.R.R. thanks the College of the Holy Cross for support. The \textit{XMM-Newton} satellite is an ESA science mission (with instruments and contributions from NASA and ESA member states). This work made use of the NASA/IPAC Extragalactic Database. J.C. would like to thank Jeremy Sanders for help with the \textsc{veusz} plotting package.

\section*{References}
Abramowicz M. A., Czerny B., Lasota J. P., Szuszkiewicz E., 1988, ApJ, 332, 646\\
Antonucci R.R.J., Miller J.S., 1985, ApJ, 297, 621\\
Arnaud K.A., 1996, ADASS 5, 17A\\
Arnaud K.A. et al., 1985, MNRAS, 217, 105\\
Ballantyne D.R., Iwasawa K., Fabian A.C., 2001, MNRAS, 323, 506\\
Ballantyne D.R., Ross R.R., Fabian A.C., 2001, MNRAS, 327, 10\\
Ballantyne D.R., Vaughan S., Fabian A.C., 2003, MNRAS, 342, 239\\
Bardeen J.M., Petterson J.A., 1975, ApJ, 195, L65\\
Boroson T.A., 2002, ApJ, 565, 78\\
Crummy J., Fabian A.C., Brandt W.N., Boller Th., 2005, MNRAS, 361, 1197\\
Czerny B., Niko\l ajuk M., R\'{o}\.{z}a\'{n}ska A., Dumont A.-M., Loska Z., \.{Z}ycki P.T., 2003, A\&A, 412, 317\\
Dasgupta S., Rao A.R., Dewangan G.C., Agrawal V.K., 2005, ApJ, 618, L87\\
Dickey J.M., Lockman F.J., 1990, ARA\&A, 28, 215\\
Fabian A.C., 1979, Proc. R. Soc. London, Ser. A, 366, 449\\
Fabian A.C., Miniutti G., 2005, astro-ph:0507409\\
Fabian A.C., Ballantyne D.R., Merloni A., Vaughan S., Iwasawa K., Boller Th., 2002, MNRAS, 331, L35\\
Fabian A.C., Rees M.J., Stella L., White N.E., 1989, MNRAS, 238, 729\\
Gallo L.C., Boller Th., Brandt W.N., Fabian A.C., Grupe D., 2004a, MNRAS, 352, 744\\
Gallo L.C., Boller Th., Brandt W.N., Fabian A.C., Vaughan S., 2004b, MNRAS, 355, 330\\
Gierli\'{n}ski M., Done C., 2004, MNRAS, 349, L7\\
Ghisellini G., Haardt F., Matt G., 2004, A\&A, 413, 535\\
Kirsch M.G.F., et al., 2004, Proc SPIE, 5488, 103\\
Krolik J.H., Hawley J.F., 2002, ApJ, 573, 754\\
Laor A., 1991, ApJ, 376, 90\\
McKernan B., Yaqoob T., Reynolds C.S., ApJ, 617, 232\\
Mineshige S., Kawaguchi T., Takeuchi M., Hayashida K., 2000, PASJ, 52, 499\\
Miniutti G., Fabian A.C., 2004, MNRAS, 349, 1435\\
Miniutti G., Fabian A.C., Goyder R., Lasenby A.N., 2003, MNRAS, 344, 22\\
Nandra K., George I.M., Mushotzky R.F., Turner T.J., Yaqoob T., 1997, ApJ, 477, 602\\
Natarajan P., Armitage P.J., 1999, MNRAS, 300, 961\\
Porquet D., Reeves J.N., O'Brien P., Brinkmann W., 2004, A\&A, 422, 85\\
Pounds K.A., Reeves J.N., King A.R., Page K.L., O'Brien P.T., 2003, MNRAS, 345, 705\\
Ross R.R., Fabian A.C., 1993, MNRAS, 261, 74\\
Ross R.R., Fabian A.C., 2005, MNRAS, 358, 211\\
Shakura N.I., Sunyaev R.A., 1973, A\&A, 24, 337\\
Sobolewska M., Done C., 2004, astro-ph:0412513\\
Streblyanska A., Hasinger G., Finoguenov A., Barcons X., Mateos S., Fabian A.C., 2005, A\&A, 432, 395\\
Tanaka Y., et al., 1995, Nature, 375, 659\\
Tanaka Y., Boller Th., Gallo L., 2005, Growing Black Holes: Accretion in a Cosmological Context, 290\\
Vaughan S., Fabian A.C., 2003, MNRAS, 348, 1415\\
Vaughan S., Boller Th., Fabian A.C., Ballantyne D.R., Brandt W.N., Tr\"{u}mper J., 2002, MNRAS, 337, 247\\
Volonteri M., Madau P., Quataert E., Rees M.J., 2005, ApJ, 620, 69\\
Walter R., Fink H.H., 1993, A\&A, 274, 105\\
Wang T., Lu Y., 2001, A\&A, 377, 52\\
Woo J.-H., Urry C.M., 2002, ApJ, 581, L5\\
\end{document}